\documentstyle[12pt]{article}

\makeatletter

\newif\ifnoncomplete

\def\final{\noncompletefalse\typeout{** FINAL form (substyle:Takashi)}}
\noncompletefalse

\newif\ifrefphysrev

\refphysrevfalse

\def \vol(#1,#2,#3){\ifrefphysrev{{\bf {#1}},
{#3} (19{#2})}\else{{{\bf {#1}}(19{#2}){#3}}}\fi}
\def \NP(#1,#2,#3){Nucl.\ Phys.\          \vol(#1,#2,#3)}
\def \PL(#1,#2,#3){Phys.\ Lett.\          \vol(#1,#2,#3)}
\def \PRL(#1,#2,#3){Phys.\ Rev.\ Lett.\   \vol(#1,#2,#3)}
\def \PRp(#1,#2,#3){Phys.\ Rep.\          \vol(#1,#2,#3)}
\def \PR(#1,#2,#3){Phys.\ Rev.\           \vol(#1,#2,#3)}
\def \PTP(#1,#2,#3){Prog.\ Theor.\ Phys.\ \vol(#1,#2,#3)}
\def \ibid(#1,#2,#3){{\it ibid.}\         \vol(#1,#2,#3)}
\def \PPNP(#1,#2,#3){Prog.\ in Part.\ and Nucl. \ Phys. \ \vol(#1,#2,#3)}

\def\thebibliography#1{
\section*{References\@mkboth
  {REFERENCES}{REFERENCES}}\list
  {[\arabic{enumi}]}{\setlength\labelwidth{2ex}
   \setlength\labelsep{0.05in} 
   \setlength\leftmargin{0.25in}
    \setlength\itemsep{0pt}
    \setlength\parsep{0pt}
   \itemsep\parskip
    \usecounter{enumi}}
    \def\newblock{\hskip .11em plus .33em minus -.22em}
    \sloppy
    \sfcode`\.=1000\relax}

\def\@bibitem#1{\item\if@filesw \immediate\write\@auxout
       {\string\bibcite{#1}{\the\c@enumi}}\fi\ignorespaces
       {\ifnoncomplete\reversemarginpar{\hspace*{-1.05in}\makebox[1in][l]
       {{\footnotesize{\sl [#1]}}}}\fi}%
       }

\def\@cite#1#2{\unskip\nobreak\relax
    {[#1]}} 

\def\citenum#1{{\def\@cite##1##2{##1}\cite{#1}}}
\def\citea#1{\@cite{#1}{}}

\newcount\@tempcntc
\def\@citex[#1]#2{\if@filesw\immediate\write\@auxout{\string\citation{#2}}\fi
  \@tempcnta\z@\@tempcntb\m@ne\def\@citea{}\@cite{\@for\@citeb:=#2\do
    {\@ifundefined
       {b@\@citeb}{\@citeo\@tempcntb\m@ne\@citea\def\@citea{,}{\bf ?}\@warning
       {Citation `\@citeb' on page \thepage \space undefined}}%
    {\setbox\z@\hbox{\global\@tempcntc0\csname b@\@citeb\endcsname\relax}%
     \ifnum\@tempcntc=\z@ \@citeo\@tempcntb\m@ne
       \@citea\def\@citea{,}\hbox{\csname b@\@citeb\endcsname}%
     \else
      \advance\@tempcntb\@ne
      \ifnum\@tempcntb=\@tempcntc
      \else\advance\@tempcntb\m@ne\@citeo
      \@tempcnta\@tempcntc\@tempcntb\@tempcntc\fi\fi}}\@citeo}{#1}}
\def\@citeo{\ifnum\@tempcnta>\@tempcntb\else\@citea\def\@citea{,}%
  \ifnum\@tempcnta=\@tempcntb\the\@tempcnta\else
   {\advance\@tempcnta\@ne\ifnum\@tempcnta=\@tempcntb \else \def\@citea{--}\fi
    \advance\@tempcnta\m@ne\the\@tempcnta\@citea\the\@tempcntb}\fi\fi}

\def\affiliation#1{\cr
\makebox[0in]{\parbox{8in}{\begin{center} {\sl #1}\end{center}}} \cr}
\def\@affiliation{}

\def\and{\cr \makebox[0in]{\rule[-1cm]{0mm}{1cm}and } \cr}

\def\maketitle{\par
 \begingroup
 \def\thefootnote{\fnsymbol{footnote}}
 \def\@makefnmark{\hbox
 to 0pt{$^{\@thefnmark}$\hss}}
 \if@twocolumn
 \twocolumn[\@maketitle]
 \else \newpage
 \global\@topnum\z@ \@maketitle \fi\thispagestyle{plain}\@thanks
 \endgroup
 \setcounter{footnote}{0}
 \let\maketitle\relax
 \let\@maketitle\relax
 \gdef\@thanks{}\gdef\@author{}\gdef\@title{}
 \gdef\@affiliation{} \let\affiliation\relax	%
 \let\thanks\relax}

\def\@maketitle{\newpage
 \null
 \vskip 0em plus 2em minus 0em     
 \ifx\@date\@empty\else
   \begin{flushright}
    {\ifnoncomplete(\today)
     \else{{\normalsize \@date}\\}\fi}      
   \end{flushright}
   \vskip 3em plus 2em minus 2em   
 \fi
 \begin{center}
  { \Large \@title \par}     
  \vskip 3em plus 1em minus 1.5em  
  {
   \lineskip .5em plus 0em minus .3em   
   \begin{tabular}[t]{c}\@author
   \end{tabular}\par}
\end{center}
 \par
 \vskip 6em plus 2em minus 4em}     

\def\abstract{\if@twocolumn
\section*{Abstract}
\else \normalsize
\fi}

\def\endabstract{\if@twocolumn\fi\par\clearpage}

\def\section{\@startsection {section}{1}{\z@}{3.5ex plus 1ex minus
    .2ex}{2.3ex plus .2ex}{\normalsize\bf}}
\def\subsection#1{\subsectioncom{\sc{#1}}}
\def\subsectioncom{\@startsection{subsection}{2}{\z@}
    {3.25ex plus 1ex minus .2ex}{1.5ex plus .2ex}{\small}}
\def\subsubsection{\@startsection{subsubsection}{3}{\z@}{3.25ex plus
1ex minus .2ex}{1.5ex plus .2ex}{\small}}

\def\@addmarginpar{\@next\@marbox\@currlist{\@cons\@freelist\@marbox
    \@cons\@freelist\@currbox}\@latexbug\@tempcnta\@ne
    \if@twocolumn
        \if@firstcolumn \@tempcnta\m@ne \fi
    \else
      \if@mparswitch
         \ifodd\c@page \else\@tempcnta\m@ne \fi
      \fi
      \if@reversemargin \@tempcnta -\@tempcnta \fi
    \fi
    \ifnum\@tempcnta <\z@  \global\setbox\@marbox\box\@currbox \fi
    \@tempdima\@mparbottom \advance\@tempdima -\@pageht
       \advance\@tempdima\ht\@marbox \ifdim\@tempdima >\z@
      \else\@tempdima\z@ \fi
    \global\@mparbottom\@pageht \global\advance\@mparbottom\@tempdima
       \global\advance\@mparbottom\dp\@marbox
       \global\advance\@mparbottom\marginparpush
    \advance\@tempdima -\ht\@marbox
    \global\ht\@marbox\z@ \global\dp\@marbox\z@
    \vskip -\@pagedp \vskip\@tempdima\nointerlineskip
    \hbox to\columnwidth
      {\ifnum \@tempcnta >\z@
          \hskip\columnwidth \hskip\marginparsep
        \else \hskip -\marginparsep \hskip -\marginparwidth \fi
       \box\@marbox \hss}
    \vskip -\@tempdima
    \nointerlineskip
    \hbox{\vrule \@height\z@ \@width\z@ \@depth\@pagedp}}

\def\ref#1{
    \@ifundefined{r@#1}{{#1}\@warning{Reference `#1'
    on page \thepage \space
    undefined}}{\edef\@tempa{\@nameuse{r@#1}}\expandafter
    \@car\@tempa \@nil\null}}
\def\refn#1{\@ifundefined{r@#1}{{#1}\@warning{Reference `#1'
    on page \thepage \space
    undefined}}{\edef\@tempa{\@nameuse{r@#1}}\expandafter
    \@car\@tempa \@nil\null}}

\def\endequationl{\eqno \@eqnnum 
$$\global\@ignoretrue}

\def\eqnarray{\stepcounter{equation}\let\@currentlabel=\theequation
\global\@eqnswtrue
\global\@eqcnt\z@\tabskip\@centering\let\\=\@eqncr
$$\arraycolsep\z@
\halign to \displaywidth\bgroup\@eqnsel\hskip\@centering
  $\displaystyle\tabskip\z@{##}$&\global\@eqcnt\@ne
  \hskip 2\arraycolsep \hfil$\displaystyle{{}##{}}$\hfil
  &\global\@eqcnt\tw@ \hskip 2\arraycolsep
  $\displaystyle\tabskip\z@{##}$\hfil
   \tabskip\@centering&\llap{##}\tabskip\z@\cr}
\def\mmodetrue{\mmode=\iftrue}

\def\eqnarrayl#1{\stepcounter{equation}\let\@currentlabel=\theequation
\label {#1}
\global\@eqnswtrue
\global\@eqcnt\z@\tabskip\@centering\let\\=\@eqncr
$$\arraycolsep\z@
\halign to \displaywidth\bgroup\@eqnsel\hskip\@centering
  $\displaystyle\tabskip\z@{##}$&\global\@eqcnt\@ne
  \hskip 2\arraycolsep \hfil$\displaystyle{{}##{}}$\hfil
  &\global\@eqcnt\tw@ \hskip 2\arraycolsep
  $\displaystyle\tabskip\z@{##}$\hfil
   \tabskip\@centering&\llap{##}\tabskip\z@\cr}
\def\label#1{
\@bsphack\if@filesw {
{\ifnoncomplete{\makebox[1in][r]{\footnotesize{\sl [#1]}}}\fi}%
\let\thepage\relax
   \xdef\@gtempa{\write\@auxout{\string
      \newlabel{#1}{{\@currentlabel}{\thepage}}}}
}\@gtempa
   \if@nobreak \ifvmode\nobreak\fi\fi\fi\@esphack}

\def\newlabel#1#2{
\@ifundefined{r@#1}{}{\@warning{Label `#1' multiply
   defined}}\global\@namedef{r@#1}{#2}}

\def\endeqnarrayl{\@@eqncr\egroup
      \global\advance\c@equation\m@ne$$\global\@ignoretrue}

\newif\if@numbersec \@numbersectrue
\def\appendix{\par\clearpage
  \setcounter{section}{0}
  \setcounter{subsection}{0}
  \def\thesection{\Alph{section}}
  \def\thesubsection{\arabic{subsection}}
  \@ifstar{\def\@sectname{Appendix}\@numbersecfalse}
          {\def\@sectname{Appendix~}\@numbersectrue}}

\def\thefigures#1{\par\clearpage\section*{Figures\@mkboth
  {FIGURES}{FIGURES}}\list
  {Fig.~\arabic{enumi}.}{\labelwidth\parindent\advance\labelwidth -\labelsep
      \leftmargin\parindent\usecounter{enumi}}}

\def\thetables#1{\par\clearpage\section*{Tables\@mkboth
  {TABLES}{TABLES}}\list
  {Table~\arabic{enumi}.}{\labelwidth-\labelsep
      \leftmargin0pt\usecounter{enumi}}}

\def\@sect#1#2#3#4#5#6[#7]#8{\ifnum #2>\c@secnumdepth
     \def\@svsec{}\else
     \refstepcounter{#1}\edef\@svsec{\ifnum #2=1 \@sectname
         \if@numbersec\csname the#1\endcsname\fi.\else
         \csname the#1\endcsname.\fi
        \hskip 1em }\fi
     \@tempskipa #5\relax
      \ifdim \@tempskipa>\z@
        \begingroup #6\relax
          \@hangfrom{\hskip #3\relax\@svsec}{\interlinepenalty \@M #8\par}
        \endgroup
       \csname #1mark\endcsname{#7}\addcontentsline
         {toc}{#1}{\ifnum #2>\c@secnumdepth \else
                      \protect\numberline{\csname the#1\endcsname}\fi
                    #7}\else
        \def\@svsechd{#6\hskip #3\@svsec #8\csname #1mark\endcsname
                      {#7}\addcontentsline
                           {toc}{#1}{\ifnum #2>\c@secnumdepth \else
                             \protect\numberline{\csname the#1\endcsname}\fi
                       #7}}\fi
     \@xsect{#5}}

\def\@sectname{}

 \def\thefootnote{\fnsymbol{footnote}}

\newcount\ieq
\newcount\jeq
\newcount\keq
\def \eq{
\multiply\ieq by 2
\jeq=\ieq
\divide\jeq by 4
\multiply\jeq by 4
\ifnum\ieq=\jeq \end{eqnarray} \keq=1 
\else
\keq=2 \begin{eqnarray} \fi
\ieq=\keq
}

\def \@magscale#1{ scaled \magstep #1}
\def \bracket<#1>{\mbox{$\langle {#1}\rangle$}}
\def \brav #1|{\mbox{$\langle {#1}|$}}
\def \cg(#1,#2,#3,#4,#5,#6){\mbox{$(#1,#2,#3,#4|#5,#6)$}}
\def \comm[#1,#2]{\mbox{$\left[{#1},{#2}\right]$}}
\def \ddt #1{\ifmmode{{\partial #1\over\partial t}} \else{${\partial
#1\over\partial t}$}\fi}

\def \dotp(#1.#2){\mbox{$(#1\cdot #2)$}}

\def \ketv #1>{\mbox{$|{#1}\rangle$}}
\def \mate<#1|#2|#3>{\mbox{$\langle {#1}|\,{#2}\,|{#3}\rangle$}}
\def \mated<#1||#2||#3>{\mbox{$\langle {#1}||\,{#2}\,||{#3}\rangle$}}
\def \mvec #1{\mbox{\boldmath{${#1}$}}}
\def \rtov(#1/#2){{\ifmmode{{\sqrt{\frac{#1}{#2}}}}
\else{{$\sqrt{\frac{#1}{#2}}$}}\fi}}
\def \sixj(#1,#2,#3,#4,#5,#6){\mbox{$\left\{\matrix
{#1&#2&#3\cr#4&#5&#6\cr}\right\}$}}
\def \Trace[#1]{\mbox{${\hbox{Tr} \left\{#1\right\}}$}}
\def \xp(#1.#2){\mbox{${(#1\times #2)}$}}

\def \etal{{\it et al.}}

\def \ie{{\it i.e.}}

\def \rhat{\ifmmode{\hat{\mvec r}}\else{$\hat{\mvec r}$}\fi}

\def \ddt #1{\ifmmode{{\partial #1\over\partial t}} \else{${\partial
#1\over\partial t}$}\fi}
\def \doublet(#1,#2){{\mbox{$\left( {\matrix{#1\cr#2\cr}} \right) $}}}
\def \triplet(#1,#2,#3){{\left(\matrix{#1\cr#2\cr#3\cr}\right)}}
\def \sixj(#1,#2,#3,#4,#5,#6)
{{\mbox{$\left\{\matrix{#1&#2&#3\cr#4&#5&#6\cr}\right\}$}}}

\def \ninej(#1,#2,#3,#4,#5,#6,#7,#8,#9)
    {\mbox{$\left\{\matrix {#1&#2&#3\cr#4&#5&#6\cr#7&#8&#9\cr}\right\}$}}

\input{epsf}

\begin{document}
\final
\date{TIT/HEP-279/NP}
\title{Direct quark transition potential for $\Lambda N \to NN$ decay }
\author{
Takashi Inoue\thanks{e-mail: tinoue@th.phys.titech.ac.jp},
Sachiko Takeuchi$^{(a)}$ and Makoto Oka \\
{\sl Department of Physics, Tokyo Institute of Technology}\\
{\sl Meguro, Tokyo 152, Japan}  \\
and \\
{\sl $^{(a)}$Department of Public Health and Environmental Science}\\
{\sl Tokyo Medical and Dental University}\\
{\sl Yushima, Bunkyo, Tokyo, 113, Japan} }
\maketitle
\abstract{
The weak $\Lambda N\to NN$ transition is studied in the valence
quark model approach.
The quark component of the two baryon system is described in the quark cluster
model and the weak transition potential is calculated by evaluating the matrix
elements of the $\Delta S=1$ effective weak Hamiltonian.
The transition potential is applied to the decay of hypernuclei and
the results are compared with available experimental data.
The results indicate that direct quark process
is significant and qualitatively different when
compared with those in conventional meson-exchange calculations.
The direct quark mechanism predicts the
violation of the $\Delta I = 1/2$ rule for this transition.
         }
\section{Introduction}
Non-leptonic weak decays of hyperons have been of interest
for many years.
Especially mesonic decays, such as $\Lambda \to N\pi$, are studied
in order to reveal the properties of the
low energy weak interactions among quarks.
Experimental data for such decays indicate a strong $\Delta I= 1/2$
enhancement compared to $\Delta I = 3/2$ component.
This $\Delta I= 1/2$ enhancement, known as the $\Delta I=1/2$
rule, is not expected naively in the standard model of the
weak interaction, and therefore its origin should be attributed for
corrections to the weak vertex due to the strong interaction.
Part of the strong corrections can be estimated by using the
renormalization group improved perturbation theory of
QCD{\cite{VSZ,GW,Paschos,Okun}}, while contributions of the low-energy hadronic
interactions are not quantitatively understood.

In order to study the corrections to the weak interaction due  to the
low-energy hadronic interaction, it may be useful to look into a new
type of weak processes, such as $NN \to NN$, and $\Lambda N \to NN$,
\ie, the two-body weak scattering processes.
It is known that the $\Lambda N \to NN$ transition plays a dominant
role in the nonmesonic decays of hypernuclei, whose data have been
accumulated in recent hypernuclear experiments.
Therefore it seems timely to study the $\Lambda N \to NN$ weak
transition from the standard theory point of view.

\par
The purpose of this paper is to study the roles of quark structure
in the two-body $\Lambda N \to NN$ weak
transition, and to construct the induced transition potential
 ~\cite{CHK,Shmatikov,OkaInoue,InoueOka}.
Because the $\Lambda N \to NN$ decay has a large momentum
transfer of approximately 420 MeV/c
(assuming the relative momentum of the initial $\Lambda$ and $N$ is zero),
the short distance dynamics of two baryons must be significant.
We here propose that the  $\Lambda N \to NN$ transition at short distance
is described by a direct quark mechanism, where a contact four-quark
interaction between the constituent quarks of baryons causes the
transition without exchanging mesons.
The four-quark vertex is taken from the low energy effective
weak Hamiltonian in the standard theory.
It contains both the $\Delta I= 1/2$ and $\Delta I= 3/2$ components.
We will see that no $\Delta I =1/2$ enhancement is seen
in the two-baryon process,
$\Lambda N \to NN$,  and thus the
$\Delta I = 3/2$ component of the weak interaction may be observed.

\begin{figure}[tp]
   \epsfysize = 50 mm
   \centerline{  \epsfbox{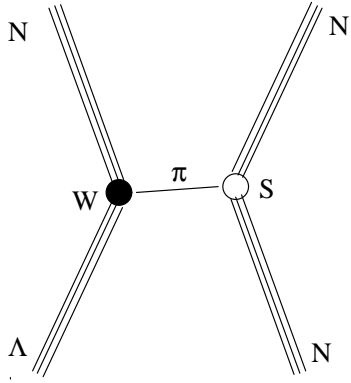} }
\caption{The diagram for the one pion exchange mechanism}
\label{fig:fig1}
\end{figure}

We will compare the direct quark transition potential with the
transition potentials based on the meson exchange mechanism
(Fig.~\ref{fig:fig1}), such as $\pi$, $K$, $\rho$ exchanges
{}~\cite{TTB,Ramos,BMZ,Choen}.
There the meson-baryon-baryon (such as $\pi-\Lambda -N$) weak
vertex is determined phenomenologically so as to describe the free hyperon
decays,
and therefore satisfies the $\Delta I= 1/2$ rule automatically.
Recent experimental data, however, have
revealed some difficulties in the meson exchange picture.
For instance, the predicted $n$-$p$ ratio is much smaller
than the experimental data for light hypernuclei.

\par
This paper is organized as follows.
In section 2,  we present the effective weak Hamiltonian
derived in the standard theory with perturbation in QCD.
In section 3, the weak Hamiltonian is applied to the
$\Lambda N to NN$ transition and the direct quark induced transition
potential $V(k,k')$ is calculated.
In section 4,  we present the explicit form of the transition
potential given in the momentum space.
In sections 5 and 6,  the transition potential is applied to the decay
of light hypernuclei and
the results are compared with available experimental data.
Discussions and conclusions are given in section 7.
\section{Effective 4-quark weak interaction}
The standard model describes hadronic weak interactions by exchanges
of the weak gauge bosons between quarks.
Because the gauge bosons are very heavy, low energy phenomena can
effectively be represented by a Hamiltonian composed of four-quark
vertices, which contain the QCD corrections on the pure weak vertex.
Such an effective Hamiltonian has been studied by several authors
{\cite{VSZ,GW,Paschos,Okun}}.
It can be computed by evaluating the perturbative QCD corrections,
using the operator product expansion and the renormalization group
equation for the Wilson coefficients.
\par
In the standard model, strangeness changing weak decay
is described by the vertex
\eq
H_W(x)=\frac{g}{2\sqrt2}J^+_{\mu}(x)W^-_{\mu}(x)+H.c. \ ,
\eq
where $W_{\mu}^{\pm}$ is the charged-W-boson field and
$J^{\pm}_{\mu}$ is the hadronic charged weak current.
Note that the standard theory contains no flavor changing neutral current.
The effective Hamiltonian for $\Delta S=\pm 1$
nonleptonic processes is defined by
\eq
\mate<|H_{eff}^{\Delta S=1}|> =
                -\frac{i}{2}\int d^4x \mate<|T H_W(x)H_W(0)|>
\eq
where RHS is the weak transition matrix element between
low-momentum hadron states composed of light quarks and differing in
strangeness by one.
We separate the mass-scale dependent coefficients and
the four-quark operators using the operator-product-expansion.
At the mass scale $\mu = M_W$ the QCD running coupling
constant $\alpha_s$ is so small that
the coefficients can be expanded perturbatively in $\alpha_s$.
Paschos \etal~ {\cite{Paschos}} takes the following Hamiltonian at
$\mu=M_W$,
\eq
      \left. H_{eff}^{\Delta S=1}\right|_{\mu=M_W}
      =  {G_f\over \sqrt{2}}
            \left[
            \xi_u
           (\bar s_{\alpha}u_{\alpha})_{V-A}
           (\bar u_{\beta}d_{\beta})_{V-A}
            + \xi_c
           (\bar s_{\alpha}c_{\alpha})_{V-A}
           (\bar c_{\beta}d_{\beta})_{V-A}
            \right]
             + H_{peng}(\mbox{t-quark})
\label{eqn:start}
\eq
where $\alpha$ and $\beta$ stand for color indices of the quark field.
The first term is a pure weak interaction at low-momentum transfer
with  $\xi_q=V_{qd}V_{qs}^*$ where matrix $V$ is
the Cabibbo-Kobayashi-Maskawa matrix.
This term contains $\Delta I = 3/2$ component as well as
$\Delta I = 1/2$ component.
The second term, $H_{peng}(\mbox{t-quark})$,
represents the first order QCD correction that is produced by a
so-called penguin diagram with the top quark in the intermediate line.
This term is needed because the top quark is heavier than the W-boson.
Starting from eq.(\ref{eqn:start}) at $\mu=M_W$
the effective Hamiltonian at low mass scale is computed with the help
of the renormalization group technique.
The one loop QCD corrections are taken into account.
Operator mixing takes place and
enhances the $\Delta I = 1/2$ component, while the $\Delta I = 3/2$
part is suppressed.
\par
The perturbation theory, of course, cannot be extended down to the low
energy region, where non-perturbative effects of QCD may modify the
weak vertex as well.
Here we employ the picture proposed by Bardeen \etal~{\cite{Bardeen}},
\ie, we assume that the perturbative correction is applied down to
$\mu^2 \simeq \mu_0^2$, where $\alpha(\mu_0^2)=1$
The effective four-quark Hamiltonians is calculated at $\mu_0^2$,
and is applied to the quark model calculation (or any other low energy
theory, such as the chiral effective theory).
\par
We use the following Hamiltonian in the present calculation :
\eq
  H_{eff}^{\Delta S=1}\left(\mu=\mu_0 \right)=
  -\frac{G_f}{\sqrt 2}\sum_{r=1,r\ne 4}^6K_r O_r
\label{eqn:heff}
\ ,
\eq
where
\eq\begin{array}{c|c|c|c|c}
  \quad  K_1   \quad  &   \quad  K_2   \quad
  &
  \quad  K_3   \quad  &   \quad  K_5   \quad
  &
  \quad  K_6   \quad \\
 \hline
   -0.284 &  0.009  & 0.026 & 0.004 & -0.021 \\
\end{array}
\nonumber
\eq
and
\eq
  O_1 &=& (\bar d_{\alpha}s_{\alpha})_{V-A}
          (\bar u_{\beta}u_{\beta})_{V-A}
         -(\bar u_{\alpha}s_{\alpha})_{V-A}
           (\bar d_{\beta}u_{\beta})_{V-A}
  \\
  O_2 &=& (\bar d_{\alpha}s_{\alpha})_{V-A}
           (\bar u_{\beta}u_{\beta})_{V-A}
         +(\bar u_{\alpha}s_{\alpha})_{V-A}
           (\bar d_{\beta}u_{\beta})_{V-A}\nonumber
  \\
     & &+2(\bar d_{\alpha}s_{\alpha})_{V-A}
         (\bar d_{\beta}d_{\beta})_{V-A}
        +2(\bar d_{\alpha}s_{\alpha})_{V-A}
         (\bar s_{\beta}s_{\beta})_{V-A}
  \\
  O_3 &=& 2(\bar d_{\alpha}s_{\alpha})_{V-A}
           (\bar u_{\beta}u_{\beta})_{V-A}
         +2(\bar u_{\alpha}s_{\alpha})_{V-A}
           (\bar d_{\beta}u_{\beta})_{V-A}\nonumber
  \\
     & &-(\bar d_{\alpha}s_{\alpha})_{V-A}
         (\bar d_{\beta}d_{\beta})_{V-A}
        -(\bar d_{\alpha}s_{\alpha})_{V-A}
         (\bar s_{\beta}s_{\beta})_{V-A}
  \\
  O_5 &=& (\bar d_{\alpha}s_{\alpha})_{V-A}
       (\bar u_{\beta}u_{\beta}+\bar d_{\beta}d_{\beta}
      + \bar s_{\beta}s_{\beta})_{V+A}
  \\
  O_6 &=& (\bar d_{\alpha}s_{\beta})_{V-A}
       (\bar u_{\beta}u_{\alpha}+\bar d_{\beta}d_{\alpha}
      + \bar s_{\beta}s_{\alpha})_{V+A}
\ .
\eq
The values of the coefficients $K_r$ are taken from ref.\cite{Paschos}.
We choose the version with the flavor dependent $\Lambda_{QCD}$,
$m_t= 200 $ GeV/c$^2$ and $\mu_0=0.24$ GeV.
The value of $\mu_0$ is chosen so as to give $\alpha_s (\mu_0^2)=1 $ for
$\Lambda_{QCD} $  = 0.1 GeV.
Among the 4-quark operators above, $O_3$ contains a part which induces the
$\Delta I=3/2$ transition, while the others are purely $\Delta I=1/2$.
One sees that the $O_1$ component is enhanced, which is purely $\Delta I=1/2$.

It is known that this $\Delta I = 1/2$ enhancement alone cannot
explain the observed ratio of $\Delta I = 1/2$ and $\Delta I = 3/2$
for the non-leptonic decays of $K$, $\Lambda$ and other strange
hadrons.
Several studies have shown that various non-perturbative effects at
low energy are crucial in understanding the large $\Delta I = 1/2$
enhancement {\cite{Bardeen,Sanda,Takizawa}}.
\section{Direct quark mechanism}
\begin{figure}[t]
   \epsfysize = 120mm
   \centerline{  \epsfbox{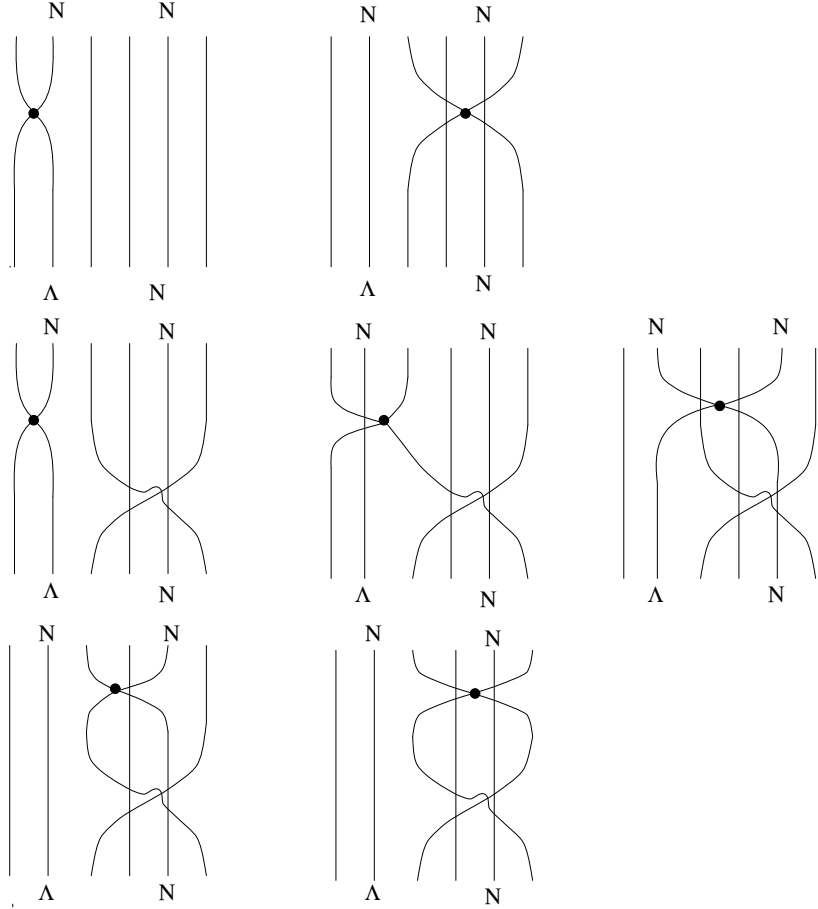} }
\caption{The diagrams for the direct quark mechanism}
\label{fig:fig2}
\end{figure}

The weak transition of two baryon systems can be described by
a transition potential $V$.
We evaluate V in the first order perturbation theory in
the effective weak Hamiltonian $H_{eff}^{\Delta S =1}$
derived in the previous section,
\eq
    V(k,k')_{{L_i,S_i,J}\atop{L_f,S_f,J}}=
      \mate<NN(k',L_f,S_f,J)|H_{eff}^{\Delta S=1}|
             \Lambda N(k,L_i,S_i,J)> .
\label{eqn:defofv}
\eq
We employ the quark cluster model
for the quark component of the two baryon systems,
\eq
\ketv BB'(L,S,J)> = {\cal A}^6 \ketv \phi(123)\phi(456)\chi(L,S,J)>
\eq
where $\phi$ is the internal wave function of the baryon
in the non-relativistic quark model
and  $\chi(\vec R)$ is the wave function for the relative
motion with $\vec R$, the relative coordinate of two baryons \cite{Oka}.
The operator ${\cal A}^6$ antisymmetrizes   the six quarks.
The transition potential eq.(\ref{eqn:defofv}) describes the direct quark
processes, in which all possible exchanges of quarks between
two baryons are included (Fig. \ref{fig:fig2}).
These direct quark processes are independent from the
meson exchange diagrams in our formulation because
the non-relativistic formalism does not allow a pair of quark and
antiquark in the intermediate state.
Therefore the full transition should be given by
a superposition of the direct quark and the meson exchange
processes.
\begin{table}[tph]
\caption{The set of transition operators}
\label{tbl:a1c11}
\eq
\begin{array}{cccccccc}
A1_{ij}
       & =
       & [ (d^{\dag}s)_i(u^{\dag}u)_j + (u^{\dag}u)_i(d^{\dag}s)_j ]
       & \otimes & 1  & \otimes & 1
       \\
A2_{ij}
       & =
       &   +
       & \otimes & \left(\vec{\sigma}_i\cdot\vec{\sigma}_j\right)
       & \otimes & 1
       \\
A3_{ij}
       & =
       &  +
       & \otimes & \left(\vec{\sigma}_i - \vec{\sigma}_j\right)
       & \otimes & (\vec{q}_{ij})
       \\
A4_{ij}
       & =
       &  +
       & \otimes &
       & \otimes & \left(\vec{P}_i - \vec{P}_j\right)
       \\
A5_{ij}
       & =
       &   -
       & \otimes &
       & \otimes & \left(\vec{P}_i + \vec{P}_j\right)
       \\
A6_{ij}
       & =
       &    +
       & \otimes & i\left(\vec{\sigma}_i \times \vec{\sigma}_j\right)
       & \otimes & (\vec{q}_{ij})
       \\
A7_{ij}
       & =
       &     +
       & \otimes &
       & \otimes & \left(\vec{P}_i - \vec{P}_j\right)
       \\
A8_{ij}
       & =
       &     -
       & \otimes &
       & \otimes & \left(\vec{P}_i + \vec{P}_j\right)
       \\
A9_{ij}
       & =
       &    -
       & \otimes & \left(\vec{\sigma}_i + \vec{\sigma}_j\right)
       & \otimes & (\vec{q}_{ij})
       \\
A10_{ij}
       & =
       &    -
       & \otimes &
       & \otimes & \left(\vec{P}_i - \vec{P}_j\right)
       \\
A11_{ij}
       & =
       &    +
       & \otimes &
       & \otimes & \left(\vec{P}_i + \vec{P}_j\right)
       \\
       \\
B1_{ij}
       & =
       & [ (d^{\dag}s)_i(d^{\dag}d)_j + (d^{\dag}d)_i(d^{\dag}s)_j ]
       & \otimes & 1
       & \otimes & 1
       \\
       \ddots \\
B11_{ij}
       & =
       &  +
       & \otimes & \left(\vec{\sigma}_i + \vec{\sigma}_j\right)
       & \otimes & \left(\vec{P}_i + \vec{P}_j\right)
       \\
       \\
C1_{ij}
       & =
       & [ (u^{\dag}s)_i(d^{\dag}u)_j + (d^{\dag}u)_i(u^{\dag}s)_j ]
       & \otimes & 1
       & \otimes & 1
       \\
       \ddots \\
C11_{ij}
       & =
       &  +
       & \otimes & \left(\vec{\sigma}_i + \vec{\sigma}_j\right)
       & \otimes & \left(\vec{P}_i + \vec{P}_j\right)
\ .
\end{array}
\nonumber
\eq
\end{table}
\par
In evaluating the transition potential we make the non-relativistic
reduction of $H_{eff}^{\Delta S=1}$ \ie \  the Breit-Fermi expansion
to first order in $p/m$.
The result is given in terms of a set of non-relativistic operators
listed in Table \ref{tbl:a1c11}.
The vectors $\vec q_{ij}$ and $\vec P_i$ are defined  by
\eq
{\vec q_{ij}}={\vec p_i}' - \vec p_i = {\vec p_j} - {\vec p_j}'
\hspace{1cm}
\vec P_i = \frac{\vec p_i + {\vec p_i}'}{2}
\eq
where $\vec p_i$ denotes the momentum of the i-th quark.
In Table \ref{tbl:a1c11}, the color operator is suppressed,
that is  unity.
Among those operators, the operators $A1,A2,B1,B2,C1$ and $C2$
are parity conserving, while the others are
of first order in $p/m$ and parity violating.
These operators are symmetric in subscripts i and j.
Appendix A gives the explicit form of $H_{eff}^{\Delta S=1}$ in terms of
$\sum_{i<j}A1_{ij} \sim \sum_{i<j}C11_{ij}$.
Because we truncate the expansion at $p/m$,
the change of the orbital angular momentum, $\Delta L$,
is restricted to 0 or $\pm 1$, namely no tensor transition is allowed.
In the present study, we restrict our initial state to $L=0$ and $1$.
Table \ref{tbl:24ch} shows 24 possible combinations of $L,S,J$, and $I$
for the initial and final states.
Note that the transition between ${}^1P_1$ and ${}^3P_1$ vanishes
because the spin change operator should change the parity as well.
\footnotesize
\begin{table}[tph]
\caption{Possible initial and final quantum numbers for
                    transitions with initial $L=0$ or $1$}
\label{tbl:24ch}
\begin{center}
\begin{tabular}{ccrcc}
  Ch&  isospin  &  spin\ orbital &  Type & W \\ \hline
  1 &   $p\Lambda\to pn$   &   ${}^{1}S_0\to{}^{1}S_0$ & 1 & 1
  \\
  2 &  &           $\to {}^{3}P_0$ & 2 & $-\sqrt{3}$
  \\
  3 &  &  ${}^{3}S_1\to {}^{3}S_1$ & 3 & 1
  \\
  4 &  &           $\to {}^{1}P_1$ & 4 & 1
  \\
  5 &  &           $\to {}^{3}P_1$ & 5 & $-\sqrt{2}$
  \\
  6 &  &  ${}^{1}P_1\to {}^{3}S_1$ & F & 1
  \\
  7 &  &           $\to {}^{1}P_1$ & G & 1
  \\
  8 &  &           $\to {}^{3}D_1$ & F & $-\sqrt{\frac52}$
  \\
  9 &  &  ${}^{3}P_0\to {}^{1}S_0$ & B & $-\sqrt{3}$
  \\
  10&  &           $\to {}^{3}P_0$ & A & 1
  \\
  11&  &  ${}^{3}P_1\to {}^{3}S_1$ & C & $-\sqrt{2}$
  \\
  12&  &           $\to {}^{3}P_1$ & A & 1
  \\
  13&  &           $\to {}^{3}D_1$ & C & $-\frac{\sqrt5}{2}$
  \\
  14&  &  ${}^{3}P_2\to {}^{3}P_2$ & A & 1
  \\
  15&  &           $\to {}^{1}D_2$ & B & $\sqrt{\frac32}$
  \\
  16&  &           $\to {}^{3}D_2$ & C & $-\frac32$
  \\
  17&   $n\Lambda\to nn$   &   ${}^{1}S_0\to{}^{1}S_0$ & 6 & 1
  \\
  18&  &           $\to {}^{3}P_0$ & 7 & $-\sqrt{3}$
  \\
  19&  &  ${}^{3}S_1\to {}^{3}P_1$ & 8 & $-\sqrt{2}$
  \\
  20&  &  ${}^{3}P_0\to {}^{1}S_0$ & I & $-\sqrt{3}$
  \\
  21&  &           $\to {}^{3}P_0$ & H & 1
  \\
  22&  &  ${}^{3}P_1\to {}^{3}P_1$ & H & 1
  \\
  23&  &  ${}^{3}P_2\to {}^{3}P_2$ & H & 1
  \\
  24&  &           $\to {}^{1}D_2$ & I & $-\sqrt{\frac32}$
\end{tabular}
\end{center}
\end{table}
\normalsize
\par
Matrix elements of
$\sum_{i<j}^6 A1_{ij} \sim \sum_{i<j}^6 C11$
are calculated for the quark cluster model wave function.
Under the condition that $\ketv \phi\phi\chi>$ is totally
antisymmetric for the exchange \{1,2,3\}$\leftrightarrow$\{4,5,6\},
the matrix element, for instance
\eq
\mate<B_3B_4|{\sum_{i<j}^6 A1_{ij}}|B_1B_2> ,
\eq
is equal to
\eq
& & \frac{6}{N}\mate<\phi\phi\chi|A1_{12}|\phi\phi\chi>
   + \frac{9}{N}\mate<\phi\phi\chi|A1_{36}|\phi\phi\chi>
  \nonumber
  \\
&-&\frac{18}{N}\mate<\phi\phi\chi|A1_{12}P_{36}|\phi\phi\chi>
   -\frac{36}{N}\mate<\phi\phi\chi|A1_{13}P_{36}|\phi\phi\chi>
  \nonumber
  \\
&-&\frac{36}{N}\mate<\phi\phi\chi|A1_{25}P_{36}|\phi\phi\chi>
   -\frac{36}{N}\mate<\phi\phi\chi|A1_{35}P_{36}|\phi\phi\chi>
  \nonumber
  \\
&-&\frac{9}{N} \mate<\phi\phi\chi|A1_{36}P_{36}|\phi\phi\chi>
  \label{eqn:seven}
\eq
where $P_{36}$ represents the permutation operator,
$3 \leftrightarrow 6$. Fig. \ref{fig:fig2} shows the diagrams corresponding
to each term of eq.(\ref{eqn:seven}).
Because each baryon wave function is totally antisymmetrized,
${\cal A}^6$ of the initial and final state can be replaced
by a single $P_{36}$ operated to the initial state.
$N$ is the normalization factor, which depends on the channel
but is nearly equal to 1 in general.
We factorize each matrix element in eq.(\ref{eqn:seven})
into the flavor-spin, orbital and color parts as
\eq
    & &
    \mate<\phi\phi\chi(k',L_f,S_f,J)|A1_{ij}(P_{36})|
         \phi\phi\chi(k,L_i,S_i,J)>
    \nonumber
    \\
    &=&
    \mate<\phi\phi\chi^{\mbox{flavor-spin}}
          |A1^{\mbox{flavor-spin}}_{ij\ z}(P_{36}^{\mbox{flavor-spin}})|
         \phi\phi\chi^{\mbox{flavor-spin}}>
    \nonumber
    \\
    & & \hspace{5mm}
    \times
    \mate<\phi\phi\chi^{\mbox{orbital}}
         |A1^{\mbox{orbital}}_{ij\ z}(P_{36}^{\mbox{orbital}})|
          \phi\phi\chi^{\mbox{orbital}}>
    \nonumber
    \\
    & & \hspace{10mm}
    \times
    \mate<\phi\phi\chi^{\mbox{color}}
         |1(P_{36}^{\mbox{color}})|
          \phi\phi\chi^{\mbox{color}}>
    \times W
    \ .
  \label{eqn:fsop}
\eq
$W$ is an algebraic factor required when
we factorize the spin and orbital matrix elements.
It is defined by
\eq
W= (-)^\lambda \sqrt{2\lambda+1} \sqrt{2J+1}
   \ninej(L_i,S_i,J,\lambda^o,\lambda^s,0,L_f,S_f,J)
   \frac{\sqrt{2L_f+1}}{\cg(L_i,\lambda^o,L_z,0,L_f,L_z)}
   \frac{\sqrt{2S_f+1}}{\cg(S_i,\lambda^s,S_z,0,S_f,S_z)}
\eq
where $\lambda^o$ and $\lambda^s$ are the ranks of the orbital and spin
operators respectively, and $\lambda = \lambda^o = \lambda^s$.
\par
In Table \ref{tbl:24ch}, we label the combinations of the initial
and the final spin-flavor part of $\ketv \phi\phi\chi>$ by ``Type''.
They are listed in Table \ref{tbl:types} in Appendix B.
In evaluating the flavor-spin matrix element,
we use the $SU(6)$ flavor-spin wave function for the nucleon and $\Lambda$.
\par
Appendix C is devoted to evaluation of the orbital matrix elements
\eq
    \mate<\phi\phi\chi^{\mbox{orbital}}(k',L_f)
         |O^{\mbox{orbital}}_{ij\ z}(P_{36}^{\mbox{orbital}})|
          \phi\phi\chi^{\mbox{orbital}}(k,L_i)>
\eq
where $O^{\mbox{orbital}}_{ij}$ is one of the following operators,
\eq
   1_{ij} \quad    &\times& \delta(\vec p'_i + \vec p'_j -\vec p_i - \vec p_j)
   \\
   \vec q_{ij} \quad
                   &\times& \delta(\vec p'_i + \vec p'_j -\vec p_i - \vec p_j)
   \\
   (\vec P_i-\vec P_j)
                   &\times& \delta(\vec p'_i + \vec p'_j -\vec p_i - \vec p_j)
   \\
   (\vec P_i+\vec P_j)
                   &\times& \delta(\vec p'_i + \vec p'_j -\vec p_i - \vec p_j)
   \ .
\eq
\par
Color matrix elements are given by
\begin{eqnarray}
\mate<\mbox{color-singlet}|1|\mbox{color-singlet}>\hspace{10mm}&=&1 , \\
\mate<\mbox{color-singlet}|1 P_{36}^{\mbox{color}}|
                                  \mbox{color-singlet}>&=&1/3 .
\end{eqnarray}
\section{Induced transition potential}
The obtained $\Lambda N \to NN $ transition potential
is written in the following form.
\eq
V(k,k')_{{L_i,S_i,J}\atop{L_f,S_f,J}} =
         -\frac{G_F}{\sqrt 2}
        \sum_{i=1}^7
        \{
        V^f_i\  f(k,k')_i
      + V^g_i\  g(k,k')_i
      + V^h_i\  h(k,k')_i \} \times W
\eq
The coefficients, $V^f$, $V^g$ and $V^h$ and
the functions, $f$, $g$ and $h$ are given
in Tables \ref{tbl:vf} $\sim$ \ref{tbl:vh12}.
The numbers are normalized in the unit of
\( 1/660 \times \sqrt 6 /1296 \)
for Type 1, 2, 3, 4, 5, A, B, C and F
and in the unit of  \( 1/660 \times \sqrt 3 /648 \)
for Type 6, 7, 8, H and I.
In Tables \ref{tbl:fkkd} $\sim$ \ref{tbl:hkkd},
$\delta(L_i,L_f)$, $I(L_i,L_f)$, $\exp[ij]$ and $F(L,X)$
are defined by
\eq
 \delta(L_i,L_f)&=& \delta_{L_i L_f} \\
 I(L_i,L_f)&=& \int Y_{L_f}^{0*}(\Omega)
                       \cos \theta Y_{L_i}^0(\Omega) d\Omega \\
 \exp[12]&=&  \frac{3\sqrt{6}}{4}
                    \exp\left[ -b^2\frac{5}{12}(k^2 + k'^2) \right]\\
 \exp[13]&=&  \frac{24\sqrt{33}}{121}
                    \exp\left[ -b^2\frac{1}{33}(7k^2 + 13k'^2) \right]\\
 \exp[25]&=&  \frac{3\sqrt{3}}{8}
                    \exp\left[ -b^2\frac{1}{6}(k^2 + k'^2) \right]\\
 \exp[35]&=&  \frac{24\sqrt{33}}{121}
                    \exp\left[ -b^2\frac{1}{33}(13k^2 + 7k'^2) \right]\\
 \exp[36]&=&  \exp\left[ -b^2\frac{1}{3}(k^2 + k'^2) \right] \\
  F(0,X)&=&   4\pi \frac{\sinh(X b^2 k k')}{X b^2 k k'} \\
  F(1,X)&=& - 4\pi \left( \frac{\sinh(X b^2 k k')}{X^2 b^4 k^2 {k'}^2}
                         -\frac{\cosh( X b^2 k k')}{X b^2 k k'} \right)
\eq
The transition potential depends on three quark model parameters that
are $m$, the constituent u, d quark mass, $m_s$, the strange quark mass
and the Gaussian parameter $b$.
The ratio $m_s/m$ is chosen as $5/3$ in the calculation of
$V^f$, $V^g$ and $V^h$.
We use $m=313$ MeV and $b=0.5$ fm in the following section.
\par
In order to study the contribution of $\Delta I = 3/2$ component of
the $H_{eff}^{\Delta S=1}$, we also calculate the transition potential
without the $\Delta I = 3/2$ component.
Table \ref{tbl:vf12} $\sim$ Table \ref{tbl:vh12} give the coefficients
when we omit the $\Delta I = 3/2$ component.
\footnotesize
%
\begin{table}[tbp]
\caption{ The coefficients $V^f$ for the full Hamiltonian.}
\label{tbl:vf}
\begin{center}
 \begin{tabular}{c|r|r|r|r|r|r|r}
   Ty.  & $V^f_1$ & $V^f_2$ & $V^f_3$ & $V^f_4$ &
 $V^f_5$ & $V^f_6$ & $V^f_7$ \\ \hline
 1&
 -77588.4  &  29011.5  &   -787.2  &   6117.8  &
 -10195.3  &  39068.5  & -34345.4  \\
 3&
 -77588.4  &  29011.5  &   -787.2  &    303.9  &
 -10195.3  &   4185.1  &    538.0  \\
 G&
  -77588.4  &  -79950.0  & -163442.3  &  -79950.0  &
  -50544.8  & -162261.5  & -162261.5  \\
 A&
 -77588.4  & -44941.4  & -34614.9  & -13581.0  &
 -22070.8  & -13756.8  & -12855.1  \\
 6&
 -77588.4  &  29011.5  &   -787.2  &    -67.5  &
 -10195.3  &   1956.7  &   2766.4  \\
 H&
 -77588.4  & -44941.4  & -34614.9  & -13581.0  &
 -22070.8  &  -9633.2  & -16978.7
 \end{tabular}
\end{center}
\end{table}
%
\begin{table}[tbp]
\caption{ The coefficients $V^g$ for the full Hamiltonian. }
\label{tbl:vg}
\begin{center}
 \begin{tabular}{c|r|r|r|r|r|r|r}
   Ty.  & $V^g_1$ & $V^g_2$ & $V^g_3$ & $V^g_4$ &
 $V^g_5$ & $V^g_6$ & $V^g_7$ \\ \hline
 2&
    -54.3  &   -297.1  &  15321.8  &   5760.4  &
    -43.6  &  21392.9  &  -2476.8  \\
 4&
    -54.3  &    -84.6  &  69818.0  &  31586.4  &
   1490.8  &  85572.9  &  84357.2  \\
 5&
     54.3  &   -122.8  & -14966.7  &  -5465.2  &
  -1767.2  &   -758.7  &  -2202.1  \\
 F&
    -54.3  &    184.6  &    314.9  &   -256.9  &
  -6683.7  & -57943.0  & -52750.3  \\
 B&
    -54.3  &    214.6  &    166.3  &  -2266.7  &
  -2239.4  & -25126.8  &  12162.8  \\
 C&
     54.3  &     45.2  &   -142.0  &     53.9  &
   -182.6  &  -1155.6  &   -261.3  \\
 7&
    -54.3  &    115.3  &  15321.8  &   5760.4  &
    656.1  &   6823.1  &  12093.1  \\
 8&
     54.3  &    152.1  & -14941.7  &  -5465.2  &
  -1467.3  &     66.0  &  -3026.8  \\
 I&
    -54.3  &    -60.3  &    216.3  &    207.4  &
  -2039.5  &  -3959.3  &  -9004.7
 \end{tabular}
\end{center}
\end{table}
%
%
\begin{table}[tbp]
\caption{ The coefficients $V^h$ for the full Hamiltonian.}
\label{tbl:vh}
\begin{center}
 \begin{tabular}{c|r|r|r|r|r|r|r}
   Ty  & $V^h_1$ & $V^h_2$ & $V^h_3$ & $V^h_4$ &
 $V^h_5$ & $V^h_6$ & $V^h_7$ \\ \hline
 2&
    -54.3  &   -297.1  &  -5116.0  &  -4602.0  &
  -1294.7  & -28090.4  &  17845.1  \\
 4&
    -54.3  &    -84.6  & -23274.8  &   -759.8  &
  -4683.2  & -43698.3  & -41266.8  \\
 5&
     54.3  &   -122.8  &   5007.7  &   -586.3  &
   4817.9  & -10152.2  &  -6002.8  \\
 F&
    -54.3  &    184.6  &      0.0  &  13521.8  &
  19700.0  & 110914.1  & 106694.0  \\
 B&
    -54.3  &    214.6  &     69.6  &   1436.8  &
   6307.2  &  21951.0  &   1772.9  \\
 C&
     54.3  &     45.2  &     95.6  &    295.2  &
    439.2  &   3666.7  &    584.1  \\
 7&
    -54.3  &    115.3  &  -5116.0  &    552.4  &
  -1469.7  &  -2249.6  &  -7995.7  \\
 8&
     54.3  &    152.1  &   4907.7  &    444.6  &
   4742.9  &  -8502.8  &  -7652.2  \\
 I&
    -54.3  &    -60.3  &   -130.4  &   1436.8  &
   6257.2  &   9305.5  &  14418.4
 \end{tabular}
\end{center}
\end{table}
\normalsize
\newcommand{\msp}{\hspace{3mm}}
\begin{table}[tbp]
\caption{The functions $f_i$ }
\label{tbl:fkkd}
\begin{center}
\begin{tabular}{l}
      ${\displaystyle
      f(k,k')_1 =
         \msp \frac{6}{N} \ ~1 \ \delta(L_i,L_f)
          \sqrt{2\pi}^3\frac{1}{b^3} \frac{1}{k^2}\delta(k'-k)  }$
     \\
       ${\displaystyle
          f(k,k')_2 =
         - \frac{18}{N} \  \frac13 \ \delta(L_i,L_f)
            \ \exp[12] \ ~F\left(L_i,\frac12 \right)   }$
     \\
       ${\displaystyle
          f(k,k')_3 =
          - \frac{36}{N} \  \frac13 \ \delta(L_i,L_f)
            \ \exp[13] \ F\left(L_i,\frac{12}{33} \right)  }$
     \\
       ${\displaystyle
         f(k,k')_4 =
           -\frac{36}{N}  \ \frac13 \  \delta(L_i,L_f)
            \ \exp[25] \msp F\left(L_i,0 \right)        }$
     \\
       ${\displaystyle
          f(k,k')_5 =
           -\frac{36}{N}  \ \frac13  \ \delta(L_i,L_f)
            \ \exp[35] \ F\left(L_i,\frac{12}{33} \right)  }$
     \\
       ${\displaystyle
          f(k,k')_6 =
            \msp \frac{9}{N}  \  ~1 \ \delta(L_i,L_f)
             \ \exp[36] \  ~F\left(L_i,\frac{2}{3} \right)  }$
     \\
       ${\displaystyle
        f(k,k')_7 =
           -\frac{9}{N}  \  \frac13 \ \delta(L_i,L_f)
            \ \exp[36]  \  ~F\left(L_i,\frac{2}{3} \right)   }$
\end{tabular}
\end{center}
\end{table}
\begin{table}[tbp]
\caption{The functions $g_i$ }
\label{tbl:gkkd}
\begin{center}
\begin{tabular}{l}
      ${\displaystyle
       g(k,k')_1 =
         \msp \frac{6}{N} \ ~1 \ I(L_i,L_f) \ \frac{k}{m}
              \sqrt{2\pi}^3\frac{1}{b^3} \frac{1}{k^2}\delta(k'-k)  }$
     \\
       ${\displaystyle
        g(k,k')_2 =
         - \frac{18}{N} \  \frac13 \ I(L_i,L_f) \ \frac{k}{m}
            \ \exp[12] \ ~F\left(L_f,\frac12 \right)   }$
     \\
       ${\displaystyle
        g(k,k')_3 =
          - \frac{36}{N} \  \frac13 \ I(L_i,L_f) \ \frac{k}{m}
            \ \exp[13] \ F\left(L_f,\frac{12}{33} \right)  }$
     \\
       ${\displaystyle
        g(k,k')_4 =
           -\frac{36}{N}  \ \frac13 \  I(L_i,L_f) \ \frac{k}{m}
            \ \exp[25] \msp F\left(L_f,0 \right)        }$
     \\
       ${\displaystyle
        g(k,k')_5 =
           -\frac{36}{N}  \ \frac13  \ I(L_i,L_f) \ \frac{k}{m}
            \ \exp[35] \ F\left(L_f,\frac{12}{33} \right)  }$
     \\
       ${\displaystyle
        g(k,k')_6 =
            \msp \frac{9}{N}  \  ~1 \ I(L_i,L_f)  \ \frac{k}{m}
             \ \exp[36] \  ~F\left(L_f,\frac{2}{3} \right)  }$
     \\
       ${\displaystyle
         g(k,k')_7 =
           -\frac{9}{N}  \  \frac13 \ I(L_i,L_f)  \ \frac{k}{m}
            \ \exp[36]  \  ~F\left(L_f,\frac{2}{3} \right)   }$
\end{tabular}
\end{center}
\end{table}
\begin{table}[tbp]
\caption{The functions $h_i$ }
\label{tbl:hkkd}
\begin{center}
\begin{tabular}{l}
      ${\displaystyle
       h(k,k')_1 =
         \msp \frac{6}{N} \ ~1 \ I(L_i,L_f) \ \frac{k'}{m}
         \sqrt{2\pi}^3\frac{1}{b^3} \frac{1}{k^2}\delta(k'-k)  }$
     \\
       ${\displaystyle
        h(k,k')_2 =
         - \frac{18}{N} \  \frac13 \ I(L_i,L_f) \ \frac{k'}{m}
            \ \exp[12] \ ~F\left(L_i,\frac12 \right)   }$
     \\
       ${\displaystyle
        h(k,k')_3 =
          - \frac{36}{N} \  \frac13 \ I(L_i,L_f) \ \frac{k'}{m}
            \ \exp[13] \ F\left(L_i,\frac{12}{33} \right)  }$
     \\
       ${\displaystyle
        h(k,k')_4 =
           -\frac{36}{N}  \ \frac13 \  I(L_i,L_f) \ \frac{k'}{m}
            \ \exp[25] \msp F\left(L_i,0 \right)        }$
     \\
       ${\displaystyle
        h(k,k')_5 =
           -\frac{36}{N}  \ \frac13  \ I(L_i,L_f) \ \frac{k'}{m}
            \ \exp[35] \ F\left(L_i,\frac{12}{33} \right)  }$
     \\
       ${\displaystyle
        h(k,k')_6 =
            \msp \frac{9}{N}  \  ~1 \ I(L_i,L_f)  \ \frac{k'}{m}
             \ \exp[36] \  ~F\left(L_i,\frac{2}{3} \right)  }$
     \\
       ${\displaystyle
        h(k,k')_7 =
           -\frac{9}{N}  \  \frac13 \ I(L_i,L_f)  \ \frac{k'}{m}
            \ \exp[36]  \  ~F\left(L_i,\frac{2}{3} \right)   }$
\end{tabular}
\end{center}
\end{table}
\footnotesize
\begin{table}[tbp]
\caption{ The coefficients $V^f$ for the $\Delta I=1/2$ Hamiltonian.}
\label{tbl:vf12}
\begin{center}
 \begin{tabular}{c|r|r|r|r|r|r|r}
   Ty.  & $V^f_1$ & $V^f_2$ & $V^f_3$ & $V^f_4$ &
 $V^f_5$ & $V^f_6$ & $V^f_7$ \\ \hline
 1&
  -77588.4  &   29011.5  &    -787.2  &    1994.3  &
  -10195.3  &   14327.3  &   -9604.2  \\
 3&
  -77588.4  &   29011.5  &    -787.2  &     303.9  &
  -10195.3  &    4185.1  &     538.0  \\
 G&
  -77588.4  &  -79950.0  & -163442.3  &  -79950.0  &
  -50544.8  & -162261.5  & -162261.5  \\
 A&
  -77588.4  &  -44941.4  &  -34614.9  &  -13581.0  &
  -22070.8  &  -11007.7  &  -15604.2  \\
 6&
  -77588.4  &   29011.5  &    -787.2  &    1994.3  &
  -10195.3  &   14327.3  &   -9604.2  \\
 H&
  -77588.4  &  -44941.4  &  -34614.9  &  -13581.0  &
  -22070.8  &  -11007.7  &  -15604.2
 \end{tabular}
\end{center}
\end{table}
%
\begin{table}[tbp]
\caption{ The coefficients $V^g$ for the $\Delta I=1/2$ Hamiltonian.}
\label{tbl:vg12}
\begin{center}
 \begin{tabular}{c|r|r|r|r|r|r|r}
   Ty.  & $V^g_1$ & $V^g_2$ & $V^g_3$ & $V^g_4$ &
 $V^g_5$ & $V^g_6$ & $V^g_7$ \\ \hline
 2&
     -54.3  &     -22.2  &   15321.8  &    5760.4  &
     422.9  &   11679.7  &    7236.5  \\
 4&
     -54.3  &     -84.6  &   69818.0  &   31586.4  &
    1490.8  &   85572.9  &   84357.2  \\
 5&
      54.3  &      60.5  &  -14950.0  &   -5465.2  &
   -1567.3  &    -208.9  &   -2751.9  \\
 F&
     -54.3  &     184.6  &     314.9  &    -256.9  &
   -6683.7  &  -57943.0  &  -52750.3  \\
 B&
     -54.3  &      31.4  &     199.6  &    -617.3  &
   -2106.2  &  -11015.1  &   -1948.9  \\
 C&
      54.3  &      45.2  &    -142.0  &      53.9  &
    -182.6  &   -1155.6  &    -261.3  \\
 7&
     -54.3  &     -22.2  &   15321.8  &    5760.4  &
     422.9  &   11679.7  &    7236.5  \\
 8&
      54.3  &      60.5  &  -14950.0  &   -5465.2  &
   -1567.3  &    -208.9  &   -2751.9  \\
 I&
     -54.3  &      31.4  &     199.6  &    -617.3  &
   -2106.2  &  -11015.1  &   -1948.9
 \end{tabular}
\end{center}
\end{table}
%
%
\begin{table}[tbp]
\caption{ The coefficients $V^h$ for the $\Delta I=1/2$ Hamiltonian.}
\label{tbl:vh12}
\begin{center}
 \begin{tabular}{c|r|r|r|r|r|r|r}
   Ty.  & $V^h_1$ & $V^h_2$ & $V^h_3$ & $V^h_4$ &
 $V^h_5$ & $V^h_6$ & $V^h_7$ \\ \hline
 2&
     -54.3  &     -22.2  &   -5116.0  &   -1165.8  &
   -1411.3  &  -10863.2  &     617.9  \\
 4&
     -54.3  &     -84.6  &  -23274.8  &    -759.8  &
   -4683.2  &  -43698.3  &  -41266.8  \\
 5&
      54.3  &      60.5  &    4941.0  &     101.0  &
    4767.9  &   -9052.6  &   -7102.4  \\
 F&
     -54.3  &     184.6  &       0.0  &   13521.8  &
   19700.0  &  110914.1  &  106694.0  \\
 B&
     -54.3  &      31.4  &     -63.7  &    1436.8  &
    6273.8  &   13520.7  &   10203.2  \\
 C&
      54.3  &      45.2  &      95.6  &     295.2  &
     439.2  &    3666.7  &     584.1  \\
 7&
     -54.3  &     -22.2  &   -5116.0  &   -1165.8  &
   -1411.3  &  -10863.2  &     617.9  \\
 8&
      54.3  &      60.5  &    4941.0  &     101.0  &
    4767.9  &   -9052.6  &   -7102.4  \\
 I&
     -54.3  &      31.4  &     -63.7  &    1436.8  &
    6273.8  &   13520.7  &   10203.2
 \end{tabular}
\end{center}
\end{table}
\normalsize
\section{Application to light hypernuclear decays}
We apply the transition potential to non-mesonic weak decays
of light hypernuclei.
We assume that the decay of $\Lambda$ in nucleus is incoherent
and that one can neglect final state interactions for two energetic
outgoing nucleons and interference effects arising from
the antisymmetrization of the final state.
Then the decay rate of a light hypernucleus is given by a sum of two-body
$\Lambda N \to NN$ transition rates.
The decay of ${}^5_{\Lambda}\mbox{He}$, for instance,
can be described in terms of the spin averaged two-body transition rates,
$ \Gamma_{\Lambda p\to pn} $ and $  \Gamma_{\Lambda n\to nn} $, as
\eq
\Gamma\left( {}^5_{\Lambda}\mbox{He} \right) =
       2\ \Gamma_{\Lambda p\to pn} + 2\  \Gamma_{\Lambda n\to nn}
\ .
\eq
where
\eq
 \Gamma =
   \frac14 \sum_{{S_i,\mu_i}\atop{S_f,\mu_f}}
   \int \! \frac{d^3 \vec K}{(2 \pi)^3} 2 \pi \delta(E.C.)
   \left|
   \int \! \frac{d^3 \vec p}{(2 \pi)^3}
   \int \! \frac{d^3 \vec q}{(2 \pi)^3}
   ~\psi_{fin}(\vec q ; \vec K)
   ~V_{{S_i,\mu_i}\atop{S_f,\mu_f}}(\vec p,\vec q)~
   \psi_{ini}(\vec p)
   \right|^2
\ .
\eq
Here $\psi_{ini}(\vec p)$ and $\psi_{fin}(\vec q ; \vec K)$
are the initial and final two-body wave functions and $V(\vec p,\vec q)$ is the
transition potential.
In the present study, we also neglect the binding energy of the initial state,
and use the following energy conservation rule,
\eq
   \delta(E.C.)= \frac{M_N}{2 K}\delta(K - K^*)
\eq
where  $K^* = 415.9$ MeV satisfies
\eq
   M_{\Lambda} + M_N  = 2 \frac{{K^*}^2}{2 M_N} + 2 M_N
\ .
\eq
Decomposing $\psi$ into partial waves and performing the $\vec K$ integration,
one obtains
\begin{equation}
 \Gamma =
   \frac{1}{(2\pi)^2} \frac{M_N K^*}{2}
   \frac14 \sum_{{L_i,S_i,J,m}\atop{L_f,S_f,J,m}}
   \left|
   \int \frac{p^2 dp}{(2 \pi)^3}
   \int \frac{q^2 dq}{(2 \pi)^3}
   \psi^{fin}_{L_f}(q ; K^*)
   ~V_{{L_i,S_i,J}\atop{L_f,S_f,J}}(p,q)~
   \psi^{ini}_{L_i}(p)
   \right|^2
\ ,
\end{equation}
where $ \psi_{L}$ is the radial part of the wave function
and $ V_{{L_i,S_i,J}\atop{L_f,S_f,J}}(p,q) $
is the partial wave decomposed transition potential.
For the decay of S-shell hypernuclei, we neglect $L\ne 0$
components of the ground state wave function and thus only the
transitions from the S wave initial states to the final plane wave
are considered.
They correspond to the channels, $1 \sim 5$ for $\Lambda p \to pn$,
and $17 \sim 19$ for $\Lambda n \to nn$, in Table \ref{tbl:24ch}.
We label the transition amplitudes  by $a$ through $f$ as
\eq
\begin{array}{cc|r|r}
  a_p \ & a_n  \ & \   {}^{1}S_0\to {}^{1}S_0 \ & \ I_f=1 \\
  b_p \ & b_n  \ &              \to {}^{3}P_0 \ &       1 \\
  c_p \ &      \ & \   {}^{3}S_1\to {}^{3}S_1 \ &       0 \\
  d_p \ &      \ &              \to {}^{3}D_1 \ &       0 \\
  e_p \ &      \ &              \to {}^{1}P_1 \ &       0 \\
  f_p \ & f_n  \ &              \to {}^{3}P_1 \ &       1
\end{array}
\nonumber
\eq
according to the widely used notation \ \cite{BD}.
Among them, the amplitudes $a, c,$ and $d$ describe
the parity conserving transitions,
while the others violate parity.
By writing the amplitudes simply as
\eq
   a_p \equiv
   \int \frac{p^2 dp}{(2 \pi)^3}
   \int \frac{q^2 dq}{(2 \pi)^3}
   \psi^{fin}_0(q ; K^*)
   V_a^{proton}(p,q)
   \psi^{ini}_0(p)
\label{eqn:radint}
\eq
for instance, one obtains
\begin{eqnarray}
 \Gamma_{\Lambda p\to pn}&=&
  \frac{M_N K^*}{2(2\pi)^2}
  \frac14
  ( |a_p|^2+|b_p|^2+3|c_p|^2+3|d_p|^2+3|e_p|^2+3|f_p|^2 ) \\
 \Gamma_{\Lambda n\to nn}&=&
  \frac{M_N K^*}{2(2\pi)^2}
  \frac14
  ( |a_n|^2+|b_n|^2+3|f_n|^2 )
\ .
\end{eqnarray}
Note that the $I=0$ states are allowed only for $\Lambda p\to pn$
while the $I=1$ final states are allowed both for
$\Lambda n\to nn$ and $\Lambda p\to pn$.
\par
We employ simple wave functions for the initial and final states.
We use the Gaussian with a short-range correlation function
for the initial state,
\eq
\psi_{ini}(\vec{R})=N_{\psi} g(\vec R)
                      \exp \left\{-\frac{1}{2B^2}\vec{R}^2\right\}
\eq
where $g$ represents the short range correlation,
\eq
g(\vec R) =  1 - C \exp\left[ - \frac{R^2}{r_0^2} \right]
\ .
\eq
For the final state we use the plane wave with the same
short range correlation function,
\eq
\psi_{fin}(\vec R;\vec K^* )= g(\vec R)
                      \exp \left\{i\vec{K^*}\cdot\vec R \right\}
\ .
\eq
We use the same $g(r)$ for the initial $\Lambda N$ and the final $NN$
only for simplicity.
We choose the parameter $B$ as $\sqrt2 \times 1.3$ fm
for the S-shell hypernuclei, which corresponds
to the shell model wave function with the Gaussian
parameter 1.3 fm for both the nucleon and $\Lambda$.
For the short range correlation we choose $ C=0.5 $
and $ r_0 = 0.5 \mbox{fm} = b$ in the present calculation.
The strength $C$ is chosen arbitrarily, while we find that the results
are qualitatively the same for other values of C except for the proton
asymmetry parameter $a_1$ (see section 6).
\par
The one-pion exchange (OPE) amplitudes are also computed for the
same wave functions.
We take the form factor $\rho$ into account,
\eq
{\cal M}=
        \int\!\!\!\int\!\!\!\int
        d^3\!\vec R d^3\!\vec r_1 d^3\!\vec r_2~
       \psi_{fin}^*(\vec R)
       \rho(\vec r_1-\frac{\vec R}{2})
       V_{\mbox{OPE}}(\vec r_1-\vec r_2)
       \rho(\vec r_2+\frac{\vec R}{2})
       \psi_{ini}(\vec R)
\ ,
\eq
where
\eq
\rho(\vec r)=\exp\left[-\frac{3}{2b^2} {\vec r}^2\right]
\eq
represents the quark density in the baryon.
Because $\Delta I=1/2$ is assumed for the weak $\Lambda N \pi$
vertex, OPE amplitudes satisfy
$a_n/a_p = b_n/b_p = f_n/f_p= \sqrt 2$,
while the $\Delta I=3/2$ amplitudes give the ratio $-{1}/{\sqrt2}$.
The weak $\Lambda N \pi$ coupling constant
is adjusted to the free $\Lambda$ decay rate.
\par
The results for the two-body transition amplitudes are
listed in Table \ref{tbl:amp}.
The numbers given under ``$\Delta I=1/2 (3/2)$'' are the results
with the pure $\Delta I=1/2 (3/2)$ transition potential.
We find that the direct quark (DQ) amplitudes are in general comparable to
the OPE ones, especially $a_p, b_p, f_p$ and $f_n$ for DQ
are larger than those for OPE.
Although $a_n$ in the full DQ is small due to the cancellation of
$\Delta I=1/2$ and $\Delta I=3/2$ amplitudes,
it seems accidental because by changing the short-range correlation factor,
the cancellation may disappear.
While OPE contains only the $\Delta I =1/2$ component,
we find large $\Delta I =3/2$ contributions for the $J=0$ DQ
transitions, $a_p, b_p, a_n$ and $b_n$.
The $\Delta I=3/2$ contributions for $f_p$ and $f_n$ are small.
The DQ amplitude $d_p$ is zero because we neglect the tensor operator
by truncating the $p/m$ expansion at the order $(p/m)$.
On the other hand, OPE has a large $d_p$ which comes from
the tensor part of the one-pion exchange interaction.
It is enhanced due to the large momentum transfer.
\begin{table}[t]
\caption{Calculated transition amplitudes in $10^{-10}$ $MeV^{-1/2}$.}
\label{tbl:amp}
\begin{center}
\small
\begin{tabular}{ccr||rrr||r}
 & & & \multicolumn{3}{c||}{Direct Quark} & OPE \\
 & isospin & spin orbital & full & $\Delta I=1/2$ & $\Delta I=3/2$ & \\
\noalign{\hrule}
$a_p$
& $p\Lambda\to pn$ & ${}^{1}S_0\to {}^{1}S_0 $
& $-78.1$ & $ -23.4$ & $-54.7$ & $  2.2$ \\
$b_p$
&                  &          $\to {}^{3}P_0 $
& $-53.5$ & $  2.0 $ & $-55.5$ & $-24.8$ \\
$c_p$
&                  & ${}^{3}S_1\to {}^{3}S_1 $
& $ -1.0$ & $ -1.0$ & $  0  $ & $  2.2$ \\
$d_p$
&                  &          $\to {}^{3}D_1 $
& $    0$ &     $0$ & $  0  $ & $-86.8$ \\
$e_p$
&                  &          $\to {}^{1}P_1 $
& $-23.2$ & $-23.2$ & $  0  $ & $-43.0$ \\
$f_p$
&                  &          $\to {}^{3}P_1 $
& $-55.4$ & $-53.8$ & $ -1.5$ & $ 20.2$ \\
\noalign{\hrule}
$a_n$
& $n\Lambda\to nn $& ${}^{1}S_0\to {}^{1}S_0 $
& $ 5.5 $ & $ -33.1$ & $ 38.6$ & $  3.1$ \\
$b_n$
&                  &          $\to {}^{3}P_0 $
& $42.2 $ & $  2.9$ & $ 39.3$ & $-35.1$ \\
$f_n$
&                  & ${}^{3}S_1\to {}^{3}P_1 $
& $-75.1$ & $-76.2$ & $  1.0$ & $ 28.6$ \\
\noalign{\hrule}
\end{tabular}
\normalsize
\end{center}
\end{table}
\section{Decays of S-shell hypernuclei }
The spin averaged  transition rates are decomposed as
\eq
   \Gamma_{\Lambda p\to pn}&=&
       \frac14\left( \Gamma_{p0} + 3\Gamma_{p1} \right)
   \ ,
   \\
   \Gamma_{\Lambda n\to nn}&=&
       \frac14\left( \Gamma_{n0} + 3\Gamma_{n1} \right)
\ ,
\eq
where $\Gamma_{NJ}$ is the transition rate for the two-body
$\Lambda N$ system with angular momentum $J$,
\eq
 \Gamma_{p0}&=&
  \frac{M_N K^*}{2(2\pi)^2}
  ( |a_p|^2+|b_p|^2 ) \\
 \Gamma_{p1}&=&
  \frac{M_N K^*}{2(2\pi)^2}
  ( |c_p|^2+|d_p|^2+|e_p|^2+|f_p|^2 ) \\
 \Gamma_{n0}&=&
  \frac{M_N K^*}{2(2\pi)^2}
  ( |a_n|^2+|b_n|^2 )  \\
 \Gamma_{n1}&=&
  \frac{M_N K^*}{2(2\pi)^2}
    |f_n|^2
\ .
\eq
Our result for $\Gamma_{NJ}$ are given in  Table \ref{tbl:lnnn}.
We find that the $J=0$ proton-induced transition rate, $\Gamma_{p0}$,
is strongly enhanced due to the $\Delta I=3/2$ transition.
Compared with the OPE result, $\Gamma_{p0}$ for DQ
is much larger and in fact is dominant while OPE is
dominated by the tensor transition included in $\Gamma_{p1}$.
This dominance of $\Gamma_{p1}$ in OPE makes the $n$-$p$ ratio,
$R_{np}$, small, where
\eq
     R_{np}  \equiv
         \frac{\Gamma_{\mbox{neutron induced}}}
              {\Gamma_{\mbox{proton induced}}}
\ .
\eq
For the spin-average hypernuclei, ${}^5_{\Lambda}\mbox{He}$, this
ratio is given by
\eq
     R_{np} =  \frac{ \Gamma_{n0} + 3 \Gamma_{n1} }
                    { \Gamma_{p0} + 3 \Gamma_{p1} }
\ .
\eq
In DQ, $\Gamma_{n1}$ is also large so that the spin averaged $R_{np}$
is as large as 1.
Thus we find that DQ and OPE predict qualitatively different values
for $R_{np}({}^5_{\Lambda}\mbox{He})$, while the decay rates,
$\Gamma({}^5_{\Lambda}\mbox{He})$, are  roughly equal.
The experimental data prefers DQ, which indicates a significant
contribution of $\Gamma_{n1}$.
\begin{table}[thpb]
\caption{Decay rates of light hypernuclei.
         All the decay rates are in the unit of  $\Gamma_{free}$.
         The experimental data for $\Gamma_{\Lambda p \to pn}$,
         $\Gamma_{\Lambda n \to nn}$, $\Gamma({}^5_{\Lambda}\mbox{He})$
         and $R_{np}({}^5_{\Lambda}\mbox{He})$ are taken from
         ref.~[18].
         Those for $R_{np}({}^4_{\Lambda}$He),
                   ${\Gamma_{n.m.}({}^4_{\Lambda}\mbox{He})}/
                          {\Gamma_{n.m.}({}^4_{\Lambda}\mbox{H}) }$
         and $\Gamma_{n0}/\Gamma_{p0}$ are taken from ref.~[19][20].
         See the main text text for the `` experimental''
         $\Gamma_{NJ}$'s. }

\label{tbl:lnnn}
\begin{center}
\newcommand{\qu}{\hspace{5mm}}
\newcommand{\qul}{\hspace{10mm}}
\small
\begin{tabular}{l||cc||c||cc||c}
              & \multicolumn{2}{c||}{Direct Quark} & OPE
              & \multicolumn{2}{c||}{DQ $\pm$ OPE} & Exp
              \\
              & {Full}     & {$\Delta I=\frac12$ } &
              & $+$ &  $-$       &     \\
\noalign{\hrule}
$\Gamma_{p0} $  &  0.177  &  0.010  & 0.012
                &  0.235  &  0.143  & $0     \sim 0.116$   \\
$\Gamma_{p1} $  &  0.071  &  0.067  & 0.193
                &  0.260  &  0.269  & $0.074 \sim 0.187$   \\
$\Gamma_{n0} $  &  0.035  &  0.021  & 0.024
                &  0.002  &  0.118  & $0.063 \sim 0.553$   \\
$\Gamma_{n1} $  &  0.111  &  0.114  & 0.016
                &  0.042  &  0.212  & $0.049 \sim 0.196$   \\
\noalign{\hrule}
$\Gamma_{\Lambda p \to pn}$
                &  0.097  &  0.053  & 0.143
                &  0.253  &  0.238  & 0.105$\pm$0.035 \\
$\Gamma_{\Lambda n \to nn}$
                &  0.092  &  0.091  & 0.018
                &  0.032  &  0.189  & 0.100$\pm$0.055 \\
$\Gamma\left( {}^5_{\Lambda}\mbox{He} \right)$
                &  0.378  &  0.295  & 0.333
                &  0.573  &  0.854  & 0.41 $\pm$ 0.14 \\
$R_{np}({}^5_{\Lambda}\mbox{He})$
                &  0.94   & 1.70    &  0.12
                &  0.12   & 0.79    &  0.93 $\pm$ 0.55\\
\noalign{\hrule}
$R_{np}({}^4_{\Lambda}$He)
                &  0.18   & 0.20    &  0.08
                &  0.004  & 0.24    &  0.18 $\pm$ 0.12 \\
$\frac{\Gamma_{n.m.}({}^4_{\Lambda}\mbox{He})}
      {\Gamma_{n.m.}({}^4_{\Lambda}\mbox{H}) }$
                &  0.63   & 0.66    &  6.58
                &  1.69   & 1.14    &  1.65 $\pm$ 0.77 \\
\noalign{\hrule}
$\Gamma_{n0}/\Gamma_{p0}$
                &  0.20   & 2.00    & 2.00
                &  0.01   & 0.854   & $-0.8 \pm 2.7$  \\
\noalign{\hrule}
$ a_1({}^5_{\Lambda}\mbox{He})$
                &  0.01   &  0.02   & $-0.19$
                &  0.20   & $-0.44$ &           \\
\noalign{\hrule}
\end{tabular}
\normalsize
\end{center}
\end{table}
\par
Recently, Schumacher proposed to check the $\Delta I =1/2$ rule
in the non-mesonic decays of the S-shell hypernuclei \ \cite{Schu,Schu2}.
He calculated the ratios of ${\Gamma_{NJ}}$
by using the following relations and the corresponding experimental data,
\eq
R_{np}({}^5_{\Lambda}He)
        &=&
           \frac{\Gamma_{n0} + 3 \Gamma_{n1}}
                {\Gamma_{p0} + 3 \Gamma_{p1}}
    \label{eqn:ratio1}
 \\
R_{np}({}^4_{\Lambda}He)
        &=&
           \frac{2 \Gamma_{n0}}
                {\Gamma_{p0} + 3 \Gamma_{p1}}
    \label{eqn:ratio2}
 \\
\frac{\Gamma({}^4_{\Lambda}\mbox{He})}
       {\Gamma({}^4_{\Lambda}\mbox{H})}
        &=&
           \frac{\Gamma_{p0} + 3 \Gamma_{p1} + 2 \Gamma_{n0}}
          {2 \Gamma_{p0} + \Gamma_{n0} + 3 \Gamma_{n1}}
    \label{eqn:ratio3}
\eq
where it is assumed that the $\Gamma_{NJ}$'s are common for
${}^5_{\Lambda}\mbox{He}$, ${}^4_{\Lambda}\mbox{He}$
and ${}^4_{\Lambda}\mbox{H}$.
These equations determine three ratios of $\Gamma_{NJ}$'s.
Using the experimental values of $\Gamma_{\Lambda p \to pn}$
and $\Gamma_{\Lambda n \to nn}$, we evaluate the ``experimental''
values of $\Gamma_{NJ}$ given in Table \ref{tbl:lnnn}.
The ratio ${\Gamma_{n0}}/{\Gamma_{p0}}$ is especially sensitive to the
$\Delta I =3/2$ mixing, \ie , it is 2 for the pure
$\Delta I=1/2$ transition, while it becomes 1/2 for the pure
$\Delta I=3/2$ transition.
The second row from the bottom in Table \ref{tbl:lnnn} gives the ratio
${\Gamma_{n0}}/{\Gamma_{p0}}$.
DQ gives a much smaller value than 2, which clearly demonstrates the
contribution of $\Delta I=3/2$.
The present data, $-0.8\pm2.7$ for the S-shell hypernuclei are not conclusive.
One also sees that DQ mechanism can reproduce the $n$-$p$ ratio
for ${}^4_{\Lambda}\mbox{He}$, and the ratio
$\Gamma({}^4_{\Lambda}\mbox{He})/\Gamma({}^4_{\Lambda}\mbox{H})$
fairly well.
\par
When the hypernucleus is polarized, the angular distribution of
the outgoing proton has an asymmetry.
It is parameterized in terms of the asymmetry parameter $a_1$  as
\eq
	W(\theta) = 1 + a_1(p) \, {\cal P}_{\Lambda} \, P_1(\cos\theta)
\ .
\eq
where  ${\cal P}_{\Lambda}$ is the polarization of $\Lambda$
and $\theta$ is the angle of the outgoing proton to the $\Lambda$ polarization.
This parameter for ${}^5_{\Lambda}\mbox{He}$ is given by \cite{BMZ}
\eq
     a_1({}^5_{\Lambda}\mbox{He})
      =  {2\sqrt{3}\,(\sqrt{2} c_p+d_p) f_p
               \over a_p^2+b_p^2+3(c_p^2+d_p^2+e_p^2+f_p^2) }
\ .
\eq
Recent experimental data indicate a large negative $a_1(p)$
for p-shell hypernuclei, $a_1(p) \leq -0.6$ \cite{Kishimoto}.
Our result for $a_1({}^5_{\Lambda}\mbox{He})$ is very small
because our $d_p$ is zero and $c_p$ is also small.
But the result is rather sensitive to the choice of the
short-range correlation, and therefore is not conclusive.
\par
So far we have not considered the interference of the DQ and OPE
amplitudes.
As we have argued in section 3, the present formalism allows us to regard
OPE independent from DQ and therefore to superpose these two amplitudes.
Because the relation between the phenomenological $\Lambda N \pi$
vertex in OPE and the effective weak Hamiltonian
$H_{eff}^{\Delta S =1}$ in DQ is not known,
the relative phase of the two amplitudes cannot be determined.
Thus we evaluate DQ $\pm$ OPE and the results are listed
in Table \ref{tbl:lnnn}.
One finds that the difference between the two choices of the relative
phase mostly appear in the neutron-induced decay rates.
$\Gamma_{nJ}$'s are suppressed in (DQ $+$ OPE) and thus the $n$-$p$
ratio $R_{np}$ becomes very small.
In this sense, the experimental data prefer the (DQ $-$ OPE)
combination.
(DQ $-$ OPE) also predicts a large negative
$a_1({}^5_{\Lambda}\mbox{He})$, which seems to agree with the
experimental value for the p-shell hypernuclei.
The ratio $\Gamma_{n0}/\Gamma_{p0}$ tends to be small $( \ll 2)$
for both ( DQ $\pm$ OPE ) and again indicates a large $\Delta I =3/2$
contribution.
In both (DQ $\pm$ OPE), we find that $\Gamma_{\Lambda p \to pn}$
is overestimated and therefore the total decay rate
$\Gamma({}^5_{\Lambda}\mbox{He})$ is too large. This is again due to
the large tensor component in $\Gamma_{p1}(\mbox{OPE})$.
It seems important that OPE is calculated in the quark interaction
point of view in order to make a reliable prediction for the
DQ - OPE interference.
\par
Recently Ramos and Bennhold studied contribution of the heavy meson
exchanges such as $K, \rho$ and $\eta$~ \cite{Ramos2}.
They indicate that such contributions are suppressed by the short
range correlation and furthermore, they tend to cancel with each other.
Recent studies of the $2\pi$ exchange mechanism
indicate that the diagrams with $\Sigma N$ and $NN$ intermediate states
cancel with each other and the net effect contributes only to the
$J=0$ amplitudes {\cite{Shmatikov2,Itonaga}}.
In all, the meson exchange contributions other than OPE
seem to be small.
Therefore one may describe the $\Lambda N \to NN$ transition well
only by the DQ and OPE.
\section{Discussions and Conclusion }
Nonmesonic decays of hypernuclei provide us with a new type of the
hadronic weak interaction.  The large momentum transfer (due to the
mass difference of $\Lambda$ and $N$) makes the transition sensitive
to the short distance quark structure of the two baryon system.
Indeed, it is found that the contribution of the direct quark processes
is as large as that of the conventional one pion exchange weak
interaction.  Furthermore, we have found that the $J=0$ transition
amplitudes show a large $\Delta I =3/2$ contribution and therefore that
the $\Delta I =1/2$ rule is significantly broken.  This may be the
first clear evidence for the $\Delta I=3/2$ weak transition, that is
expected in the standard theory of the weak interaction.

We have employed, in the present analysis, an effective weak
Hamiltonian for quarks, which takes account of the one-loop
perturbative QCD corrections.  Then we have evaluated the transition
amplitudes using the quark model wave functions of baryons to the
first order in the weak interaction.
The flavor/spin structure of the amplitudes reflects the SU(6)
symmetry of the baryon wave functions, which have been verified in the
low energy baryon spectrum and properties of the baryons.

We have derived an effective $\Lambda N \to NN$ transition potential,
and applied it to the s-shell hypernuclear decays.  It is found that
the decay amplitudes show distinctive features when they are compared
to the one-pion exchange.  Especially, the ratio of the
neutron-induced and the proton-induced decay rates is discriminative
of these mechanisms.
It is suggested that the ratios of the transition rates with various
spin-isospin specification can be obtained from the experimental data
for the s-shell hypernuclei and they are useful in testing different
mechanisms of the transition.  Further experimental studies are most
desirable.

There are a number of remaining problems.
The relation between the phenomenological $\Lambda \to N\pi $
transition Hamiltonian and the effective quark Hamiltonian is to be
studied.  It is favorable to apply the same quark Hamiltonian to the
mesonic decay as well so that a unified view of the hypernuclear decay
is obtained.  The $\Delta I =1/2$ enhancement mechanism for the
mesonic decay is especially important in this regard.  This line of
study is underway and will be reported elsewhere {\cite{Inoue2}}.

For hypernuclei other than the s-shell systems,  we need a realistic
calculation combined with the nuclear structure analysis.  We have
provided the baryonic two-body transition potential that can be used in
any hypernuclear structure calculations.  Because of the nonlocal
structure due to the quark exchange effects the transition potential
is given in the momentum space, but the transformation into the
coordinate space is straightforward.

We have not considered so far the second order process with a
$\Sigma-N$ intermediate state induced by a strong pion (meson)
and/or quark exchanges.  The weak $\Sigma N \to NN$ decay can be also
computed in the same direct quark mechanism.  It is found that the
mixing of $\Sigma N$ does not change our main conclusions mentioned
above, though its contribution is not negligible quantitatively.  The
results of this calculation will be published in a separate
article {\cite{Inoue3}}.
\appendix
\section{Non-relativistic forms of $O_1 \sim O_6 $}
The Hamiltonian $H_{eff}^{\Delta S=1}$ is given by eq.(\ref{eqn:heff}):
\eq
  H_{eff}^{\Delta S=1}\left(\mu \sim \mu_0 \right)=
  -\frac{G_f}{\sqrt 2}\sum_{r=1,r\ne 4}^6K_r O_r
\eq
The operators, $O_2 \sim O_6$, contain the terms
$(\bar d s)(\bar s s )$, which we omit because
they do not contribute in the valence quark model.
Then the Breit-Fermi expansion of $O_1 \sim O_6$ is given in terms
of the operators $A1 \sim C11$ by
\eq
O_1&=&  (A1-C1) - (A2-C2) - \frac{\delta}{2}(A3-C3)
        \nonumber
        \\
   & &    - \frac{2\mu+\delta}{2}(A4-C4) - \frac{\delta}{2}(A5-C5)
        \nonumber
        \\
   & &    - \frac{2\mu+\delta}{2}(A6-C6)
          - \frac{\delta}{2}(A7-C7) - \frac{\delta}{2}(A8-C8)
\label{eqn:o1}
        \\
O_2&=&  (A1+2B1+C1) - (A2+2B2+C2) - \frac{\delta}{2}(A3+2B3+C3)
        \nonumber
        \\
   & &    - \frac{2\mu+\delta}{2}(A4+2B4+C4) - \frac{\delta}{2}(A5+2B5+C5)
        \nonumber
        \\
   & &    - \frac{2\mu+\delta}{2}(A6+2B6+C6)
          - \frac{\delta}{2}(A7+2B7+C7) - \frac{\delta}{2}(A8+2B8+C8)
        \\
O_3&=&  (2A1-B1+2C1) - (2A2-B2+2C2) - \frac{\delta}{2}(2A3-B3+2C3)
        \nonumber
        \\
   & &    - \frac{2\mu+\delta}{2}(2A4-B4+2C4) - \frac{\delta}{2}(2A5-B5+2C5)
        \nonumber
        \\
   & &    - \frac{2\mu+\delta}{2}(2A6-B6+2C6)
          - \frac{\delta}{2}(2A7-B7+2C7) - \frac{\delta}{2}(2A8-B8+2C8)
        \\
O_5&=&  (A1+B1) + (A2+B2)
          - \frac{\delta}{2}(A6+B6)
          + \frac{\delta}{2}(A7+B7) + \frac{\delta}{2}(A8+B8)
        \nonumber
        \\
   & &    - \frac{\delta}{2}(A9+B9)
          - \frac{2\mu+\delta}{2}(A10+B10) + \frac{\delta}{2}(A11+B11)
        \\
O_6&=&-2\left\{  (B1+C1)
          - \frac{\delta}{4}(B3+C3)
          + \frac{\delta}{4}(B4+C4) + \frac{\delta}{4}(B5+C5)
        \right.  \nonumber \\
   & & \hspace{1cm}\left.  + \frac{2\mu+\delta}{4}(B9+C9)
          + \frac{\delta}{4}(B10+C10) + \frac{\delta}{4}(B11+C11)
         \right\}
\label{eqn:o6}
\eq
where
\eq
\mu\equiv\frac{m_s+m}{2m_s m}
\ \mbox{and}\
\delta\equiv\frac{m_s-m}{2m_s m}
\ ,
\eq
are given in terms of $m$, the mass of constituent $u, d$
quarks, and $m_s$, the mass of constituent $s$ quark.
In the present calculation, we use $m=313$ MeV and $m/m_s=3/5$.
Note that we have made the following Fierz transformation on $O_6$
so that the color part becomes unity.
\eq
  O_6 &=& (\bar d_{\alpha}s_{\beta})_{V-A}
           (\bar u_{\beta}u_{\alpha}+\bar d_{\beta}d_{\alpha}
      + \bar s_{\beta}s_{\alpha})_{V+A} \\
      &=& -2(\bar d_{\alpha}u_{\alpha})_{S+P}
       (\bar u_{\beta}s_{\beta})_{S-P}
          -2(\bar d_{\alpha}d_{\alpha})_{S+P}
       (\bar d_{\beta}s_{\beta})_{S-P}
\eq
\section{Flavor-spin part of $ \ketv \phi \phi \chi>$ }
\newcommand{\plplp}{\frac{1}{\sqrt2}(p\Lambda+\Lambda p) }
\newcommand{\nlpln}{\frac{1}{\sqrt2}(n\Lambda+\Lambda n) }
\newcommand{\pnpnp}{\frac{1}{\sqrt2}(pn+np) }
\newcommand{\udmdu}{\frac{1}{\sqrt2}(\uparrow\downarrow - \downarrow\uparrow) }
\newcommand{\udpdu}{\frac{1}{\sqrt2}(\uparrow\downarrow + \downarrow\uparrow) }
\newcommand{\uu}{ \uparrow\uparrow }
\footnotesize
\begin{table}[tbp]
\caption{Flavor-Spin part of $\ketv \phi\phi\chi>$. }
\label{tbl:types}
\begin{center}
\begin{tabular}{c||ccc|ccc}
      &\multicolumn{3}{c|}{Initial}& \multicolumn{3}{c}{Final}\\ \hline
   Ty. &  Orb. & Flavor & Spin & Orb. &  Flavor & Spin \\   \hline
  1 & S & $\plplp$ &  $\udmdu$ & S & $\pnpnp$ & $\udmdu$
  \\
  2 &   &  $+$     &    $-$    & P &  $+$    &    $+$
  \\
  3 &   &  $-$     &    $+$    & S &  $-$    &    $+$
  \\
  4 &   &  $-$     &    $+$    & P &  $-$    &   $-$
  \\
  5 &   &  $-$     &   $\uu$   & P &  $+$    &   $\uu$
  \\
  A & P &  $+$     &   $\uu$   & P &  $+$    &   $\uu$
  \\
  B &   &  $+$     &  $\udpdu$ & S &  $+$    &  $\udmdu$
  \\
  C &   &  $+$     &   $\uu$   & S &  $-$    &   $\uu$
  \\
  F &   &  $-$     &  $\udmdu$ & S &  $-$    &  $\udpdu$
  \\
  G &   &  $-$     &     $-$   & P &  $-$    &    $-$
  \\
  \hline
  6 & S &$\nlpln$  &  $\udmdu$ & S &  $nn$   &  $\udmdu$
  \\
  7 &   &  $+$     &    $-$    & P &  $nn$   &    $+$
  \\
  8 &   &  $-$     &   $\uu$   & P &  $nn$   &   $\uu$
  \\
  H & P &  $+$     &   $\uu$   & P &  $nn$   &   $\uu$
  \\
  I &   &  $+$     &  $\udpdu$ & S &  $nn$   &  $\udmdu$
\end{tabular}
\end{center}
\end{table}
\normalsize
Table \ref{tbl:types} gives the flavor-spin part of
$\ketv \phi\phi\chi>$ for each Type.
One sees that $\ketv \phi\phi\chi>$ is totally antisymmetric under
the exchange \{1,2,3\}$\leftrightarrow$\{4,5,6\}.
We choose either $s_z=0$ or $1$
so that  \cg(s_i,\lambda^s,s_z,0,s_f,s_z) is not zero.
These ``Type''s are refered to by the channels in Table \ref{tbl:24ch}.
\section{Orbital matrix elements}
\newcommand{\veccent}{ \frac{\vec r_1+\vec r_2+\vec r_3}{3} }
\newcommand{\vecrel}{ \textstyle  \frac{\vec r_1+\vec r_2+\vec r_3}{3}
                      - \frac{\vec r_4+\vec r_5+\vec r_6}{3} }
\begin{table}[tbp]
\caption{Definition of the Jacobi coordinates
          and their conjugate momenta}
\label{tbl:jacobi}
\begin{center}
\begin{tabular}{lcll}
      $ \vec \xi_{12}  \ \ \ = $
      $ \vec r_1 - \vec r_2    $
    & $ \leftrightarrow        $
    & $ \vec p_{12}   \ \ =    $
      $ \frac{\vec p_1 - \vec p_2}{2}   $
    \\
      $ \vec \xi_{12-3}  = $
      $ \vec r_3 - \frac{\vec r_1 - \vec r_2}{2}  $
    & $ \leftrightarrow   $
    & $ \vec p_{12-3}   = $
      $ \frac23\vec p_3 - \frac{\vec p_1 + \vec p_2}{3}  $
    \\
      $ \vec \xi_{45}   \ \ \ = $
      $ \vec r_4 - \vec r_5     $
    & $ \leftrightarrow         $
    & $ \vec p_{45}  \ \ \ =    $
      $ \frac{\vec p_4 - \vec p_5}{2}  $
    \\
      $ \vec \xi_{45-6}  = $
      $ \vec r_6 - \frac{\vec r_4 - \vec r_5}{2}  $
    & $ \leftrightarrow   $
    & $ \vec p_{45-6}  =  $
      $ \frac23\vec p_6 - \frac{\vec p_4 + \vec p_5}{3}  $
    \\
      $ \vec R    \ \ \ \ = $
      $ \vecrel             $
    & $ \leftrightarrow     $
    & $ \vec P   \ \ \ \ =  $
      $ \frac{ \vec p_1 + \vec p_2 + \vec p_3
                           - \vec p_4 - \vec p_5 - \vec p_6}{2}   $
    \\
      $ \vec R_G    \ \ \ =  $
      $ \frac{  \vec r_1 + \vec r_2 + \vec r_3
                            + \vec r_4 + \vec r_5 + \vec r_6 }{6} $
    & $ \leftrightarrow      $
    & $ \vec P_G   \ \ \ =  $
      $ {\textstyle \vec p_1+\vec p_2+\vec p_3+\vec p_4+\vec p_5+\vec p_6} $
\end{tabular}
\end{center}
\end{table}
The orbital part of the internal wave function of
$\Lambda$ and $N$ is taken as
\eq
    \phi(1,2,3)^{\mbox{orbital}} =
    N
    e^{ -\frac{1}{2b^2}(\vec{r_1}- \veccent )^2  }
    e^{ -\frac{1}{2b^2}(\vec{r_2}- \veccent )^2  }
    e^{ -\frac{1}{2b^2}(\vec{r_3}- \veccent )^2  }
\eq
where $b$ denotes the size of the baryon. We choose $b=0.5$ fm.
It is convenient to use the Jacobi coordinates,
defined in Table \ref{tbl:jacobi},
in writing the orbital wave function of the two baryon system,
\eq
    \phi(1,2,3)^{\mbox{orbital}}  \phi(4,5,6)^{\mbox{orbital}}
    \chi\left( \vecrel \right)
\ .
\eq
The function $\phi(1,2,3)^{\mbox{orbital}}$
is written in terms of the  Jacobi coordinates $\vec{\xi}_{12}$ and
$\vec{\xi}_{12-3}$ as
\eq
    \phi(1,2,3)^{\mbox{orbital}} =
    \left(\frac{1}{2\pi b^2}\right)^{\frac34}
    \left(\frac{2}{3\pi b^2}\right)^{\frac34}
    \exp\left\{ -\frac{1}{4b^2}{\vec{\xi}_{12}}^2   \right\}
    \exp\left\{ -\frac{1}{3b^2}{\vec{\xi}_{12-3}}^2 \right\}
\eq
or in the momentum space as
\eq
   \phi(1,2,3)^{\mbox{orbital}}  =
    (8\pi b^2)^{4/3}(6\pi b^2)^{3/4}
    \exp\left\{-b^2 {\vec p_{12}}^2\right\}
    \exp\left\{-\frac{3}{4b^2}{\vec p_{12-3}}^2\right\}
\ .
\eq
\par
Explicit forms of the orbital matrix elements
\eq
 \mate<~\phi\phi (2 \pi)^3 \delta( \vec P' - \vec k' ) |  {\cal O}_{ij}(P_{36})
|
             \phi\phi (2 \pi)^3 \delta( \vec P -  \vec k  )~ >
\eq
are listed in Table \ref{tbl:ome1} $\sim$ Table \ref{tbl:ome4}.
The matrix elements
\eq
    \mate<~\phi\phi\chi(k',L_f)
         |{\cal O}_{ij\ z}(P_{36})|
          \phi\phi\chi(k,L_i)~>
\eq
are given by the integration
\eq
\int \!\! d \hat k' \!\!  \int \!\! d \hat k
  \ Y_{L_f}^{0*}( \hat k')
  ~\mate<~\phi\phi (2 \pi)^3 \delta( \vec P' - \vec k' ) |  {\cal
O}_{ij}(P_{36}) |
             \phi\phi (2 \pi)^3 \delta( \vec P -  \vec k  )~>
  ~Y_{L_i}^0( \hat k)
\ .
\eq
\newcommand{\kdotkd}{\vec k \cdot \vec k'}
\begin{table}[tbp]
\caption{ Orbital matrix elements.}
\label{tbl:ome1}
\begin{center}
\small
\begin{tabular}{l|cc}
     ${\cal O}_{ij}(P_{36})$ & &
     $ \mate<~\phi\phi (2 \pi)^3 \delta( \vec P' - \vec k' ) | {\cal
O}_{ij}(P_{36}) |
             \phi\phi (2 \pi)^3 \delta( \vec P -  \vec k  )~ > $
   \\ \hline
   \\
   $1_{12}$
   &
   & ${ \displaystyle
     \sqrt{2\pi}^3\frac{1}{b^3}\frac{1}{k^2}\delta(k'-k)   }$
   \\
   $1_{12}\ P_{36}$
   &
   & ${\displaystyle
      \frac{3\sqrt 6}{4}
      \exp\left[-b^2 ( \frac{5}{12}k^2
                     + \frac12 \kdotkd
                     + \frac{5}{12}k'^2 ) \right]  }$
   \\
   $1_{13}\ P_{36}$
   &
   & ${\displaystyle
      \frac{24\sqrt{33}}{121}
      \exp\left[-b^2\frac{1}{33}(7k^2 + 12 \kdotkd + 13k'^2)\right]
     }$
   \\
   $1_{25}\ P_{36}$
   &
   & ${\displaystyle
      \frac{3\sqrt{3}}{8}
      \exp\left[-b^2\frac{1}{6}(k^2+k'^2)\right]  }$
   \\
   $ 1_{35}\ P_{36}$
   &
   & ${\displaystyle
      \frac{24\sqrt{33}}{121}
      \exp\left[-b^2\frac{1}{33}(13k^2 + 12 \kdotkd + 7k'^2)\right] }$
   \\
   $1_{36}$
   &
   & ${\displaystyle
        \exp\left[-b^2\frac{1}{3}(k^2 + 2 \kdotkd + k'^2)\right]   }$
   \\
   $1_{36}\ P_{36}$
   &
   & ${\displaystyle
     \exp\left[-b^2\frac{1}{3}(k^2 + 2 \kdotkd + k'^2)\right]  }$
\end{tabular}
\normalsize
\end{center}
\end{table}
\begin{table}[tbp]
\caption{ Orbital matrix elements.}
\label{tbl:ome2}
\begin{center}
\small
\begin{tabular}{l|cc}
     $ {\cal O}_{ij}(P_{36}) $ & &
     $ \mate<~\phi\phi (2 \pi)^3 \delta( \vec P' - \vec k' ) | {\cal
O}_{ij}(P_{36}) |
              \phi\phi (2 \pi)^3 \delta( \vec P -  \vec k  )~ > $
   \\ \hline
   \\
   $\vec q_{12}$
   &
   & 0
   \\
   $\vec q_{12}\ P_{36}$
   &
   & 0
   \\
   $\vec q_{13}\ P_{36}$
   &
   & ${\displaystyle
     \frac{1}{66}
     \left\{  -36 \vec k + 12 \vec k'   \right\}
     \mate<|1_{13}P_{36}|>  }$
   \\
   $\vec q_{25}\ P_{36}$
   &
   &${\displaystyle
     \frac{1}{66}
     \left\{     -33 \vec k + 33 \vec k'      \right\}
     \mate<|1_{25}P_{36}|>   }$
   \\
   $\vec q_{35}\ P_{36}$
   &
   &${\displaystyle
     \frac{1}{66}
     \left\{  -12 \vec k + 36 \vec k'      \right\}
     \mate<|1_{35}P_{36}|>   }$
   \\
   $\vec q_{36}$
   &
   &${\displaystyle
     \frac{1}{66}
     \left\{  -66 \vec k + 66 \vec k'      \right\}
     \mate<|1_{36}|>   }$ \hspace{6mm}
   \\
   $\vec q_{36}\ P_{36}$
   &
   &${\displaystyle
     \frac{1}{66}
     \left\{ + 22 \vec k + 22 \vec k'      \right\}
     \mate<|1_{36}P_{36}|>   }$
\end{tabular}
\normalsize
\end{center}
\end{table}
\begin{table}[tbp]
\caption{ Orbital matrix elements.}
\label{tbl:ome3}
\begin{center}
\small
\begin{tabular}{l|cc}
   $ {\cal O}_{ij}(P_{36}) $ & &
   $ \mate<~\phi\phi (2 \pi)^3 \delta( \vec P' - \vec k' ) | {\cal
O}_{ij}(P_{36}) |
            \phi\phi (2 \pi)^3 \delta( \vec P -  \vec k  )~ > $
   \\ \hline
   \\
   $ ( \vec P_1 - \vec P_2 ) $
   &
   & 0
   \\
   $ (\vec P_1 - \vec P_2 )\ P_{36}$
   &
   & 0
   \\
   $ (\vec P_1 - \vec P_3) \ P_{36}$
   &
   &${\displaystyle
    \frac{1}{66}
    \left\{   + 36 \vec k - 12 \vec k'     \right\}
    \mate<|1_{13}P_{36}|>   }$
   \\
   $ (\vec P_2 - \vec P_5) \ P_{36}$
   &
   &${\displaystyle
    \frac{1}{66}
    \left\{  + 33 \vec k + 33 \vec k'   \right\}
    \mate<|1_{25}P_{36}|>   }$
   \\
   $ (\vec P_3 - \vec P_5) \ P_{36}$
   &
   &${\displaystyle
    \frac{1}{66}
    \left\{ -12 \vec k + 36 \vec k'    \right\}
     \mate<|1_{35}P_{36}|>   }$
   \\
   $ (\vec P_3 - \vec P_6) $
   &
   &${\displaystyle
     \frac{1}{66}
     \left\{ + 22 \vec k + 22 \vec k'    \right\}
     \mate<|1_{36}|>   }$ \hspace{6mm}
   \\
   $ (\vec P_3 - \vec P_6) \ P_{36}$
   &
   &${\displaystyle
    \frac{1}{66}
    \left\{  - 66 \vec k + 66 \vec k'  \right\}
     \mate<|1_{36}P_{36}|>   }$
\end{tabular}
\normalsize
\end{center}
\end{table}
\begin{table}[tbp]
\caption{ Orbital matrix elements.}
\label{tbl:ome4}
\begin{center}
\small
\begin{tabular}{l|cc}
   $ {\cal O}_{ij}(P_{36}) $ & &
   $ \mate<~\phi\phi (2 \pi)^3 \delta( \vec P' - \vec k' ) | {\cal
O}_{ij}(P_{36}) |
            \phi\phi (2 \pi)^3 \delta( \vec P -  \vec k  )~ > $
   \\ \hline
   \\
   $ (\vec P_1 + \vec P_2) $
   &
   &${\displaystyle
    \frac{1}{66}
    \left\{ + 22 \vec k + 22 \vec k'  \right\}
    \mate<|1_{12}|> }$ \hspace{6mm}
   \\
   $ (\vec P_1 + \vec P_2) \ P_{36}$
   &
   &${\displaystyle
    \frac{1}{66}
    \left\{  + 33 \vec k - 33 \vec k'    \right\}
    \mate<|1_{12}P_{36}|>   }$
   \\
   $ (\vec P_1 + \vec P_3) \ P_{36}$
   &
   &${\displaystyle
    \frac{1}{66}
    \left\{   + 12 \vec k - 48 \vec k'   \right\}
    \mate<|1_{13}P_{36}|>   }$
   \\
   $ (\vec P_2 + \vec P_5) \ P_{36}$
   &
   & 0
   \\
   $ (\vec P_3 + \vec P_5) \ P_{36}$
   &
   &${\displaystyle
     \frac{1}{66}
     \left\{   - 48 \vec k + 12 \vec k'  \right\}
     \mate<|1_{35}P_{36}|>   }$
   \\
   $ (\vec P_3 + \vec P_6) $
   &
   & 0
   \\
   $ (\vec P_3 + \vec P_6) \ P_{36}$
   &
   & 0
 \end{tabular}
\normalsize
\end{center}
\end{table}

\newpage
\begin{thebibliography}{99}
%
\bibitem{VSZ}
A.I.~Vainshtein, V.I.~Zakharov and M.A.~Shifman,
Sov.\ Phys.\ JETP, \vol(45,77,670)


\bibitem{GW}
F.~J.~Gillman and M.~B.~Wise, \PR(D20,79,2382)

\bibitem{Paschos}
E.A.~Paschos, T.~Schneider, and Y.L.~Wu, \NP(B332,90,285)

\bibitem{Okun}
L.B.~Okun, {\sl Leptons and Quarks} ~(North Holland, The Netherlands, 1982)


\bibitem{CHK}
C.~Y.~Cheung, D.~P.~Heddle and L.~S.~Kisslinger, \PR(C27,83,335)

\bibitem{Shmatikov}
K.~Maltman and M.~Shmatikov, \PL(B311,94,1)

\bibitem{OkaInoue}
M.~Oka, T.~Inoue and S.~Takeuchi,
    Properties \& Interactions of Hyperons, \\
      Proceedings of the U.S.-Japan Seminar
       ed. by B.~F.~Gibson, P.~D.~Barnes and K.~Nakai
       ~(World Scientific, 1994)

\bibitem{InoueOka}
T.~Inoue, S.~Takeuchi and M.~Oka, \NP(A577,94,281c)


\bibitem{TTB}
K.~Takeuch, H.~Takaki and H.~Bando, \PTP(73,85,841)

\bibitem{Ramos}
A.~Ramos, E.~van ~Meijgaard, C.~Bennhold, and B.~K.~Jennings, \\
  \NP(A544,92,703)


\bibitem{BMZ}
H.~Bando, T.~Motoba and J.~Zofka, Int.\ Jour.\ Mod.\ Phys. \vol(A5,90,4021)

\bibitem{Choen}
J.~Cohen, \PPNP(25,90,139)


\bibitem{Bardeen}
W.A.~Bardeen, A.J.~Buras and J.M.~Gerard \PL(B192,87,138)


\bibitem{Sanda}
T.~Morozumi, C.S.~Lim, A.I.~Sanda, \PRL(65,90,404)

\bibitem{Takizawa}
M.~Takizawa, T.~Inoue, and M.~Oka, INS-Rep.-1089\\
to be published in Prog. Theor. Phys. Suppl.

\bibitem{Oka}
M.\ Oka and K.\ Yazaki, \PL(B90,80,41)


\bibitem{BD}
M.M.~Block and R.H.~Dalitz, \PRL(11,63,96)


\bibitem{Szymanski}
J.J.~ Szymanski \etal~ \PR(C43,91,849)


\bibitem{Schu}
R.A.~Schumacher, \NP(A547,92,143c)

\bibitem{Schu2}
R.A.~Schumacher,
    Properties \& Interactions of Hyperons, \\
      Proceedings of the U.S.-Japan Seminar
       ed. by B.~F.~Gibson, P.~D.~Barnes and K.~Nakai
       ~(World Scientific, 1994)


\bibitem{Kishimoto}
T.~Kishimoto,
    Properties \& Interactions of Hyperons, \\
      Proceedings of the U.S.-Japan Seminar
       ed. by B.~F.~Gibson, P.~D.~Barnes and K.~Nakai
       ~(World Scientific, 1994)

\bibitem{Ramos2}
A.~Ramos and C.~Bennhold, \NP(A577,94,287c)

\bibitem{Shmatikov2}
M.~Shmatikov, Preprint IAE-5708/2 M.~1994

\bibitem{Itonaga}
K.~Itonaga, T.~Ueda and T.~Motoba, \NP(A577,94,301c)

\bibitem{Inoue2}
T.~Inoue, M.~Takizawa and M.~Oka,  to be published

\bibitem{Inoue3}
T.~Inoue, S.~Takeuchi and M.~Oka,  to be published
%
\end{thebibliography}
\end{document}